\documentclass[12pt]{article}

\usepackage[inline,shortlabels]{enumitem}
\usepackage{booktabs}
\usepackage[T1]{fontenc}
\usepackage[margin=1in]{geometry}
\usepackage{graphicx}
\usepackage{caption}
\usepackage{float}
\usepackage{url}
\usepackage[english]{babel}
\usepackage{authblk}
\usepackage{amssymb,amsmath,amsthm}
\usepackage{mathtools}
\usepackage{graphicx}
\usepackage{setspace}
\usepackage{dsfont}
\usepackage{longtable}
\usepackage{color}
\usepackage{xcolor}
\usepackage{bbm}
\usepackage{multirow}
\usepackage{subcaption}
\usepackage{refcount}
\usepackage{accents}
\usepackage{thmtools}
\usepackage{changepage} 
\usepackage{xr-hyper}

\usepackage[titletoc,title]{appendix}

\usepackage[normalem]{ulem}

\DeclareMathOperator{\logit}{logit}
\usepackage[colorlinks,citecolor=blue,urlcolor=blue]{hyperref}
\usepackage[utf8]{inputenc}
\usepackage[authoryear]{natbib}
\usepackage{enumitem}
\bibpunct{(}{)}{;}{a}{}{,}

{
\theoremstyle{definition}
\newtheorem{assumption}{}
}

\newtheorem{theorem}{Theorem}

\newcommand{\titlepaper}{Optimizing Precision and Power by Machine Learning in Randomized Trials, with an Application to COVID-19}
\author[1]{Nicholas Williams}
\author[2]{Michael Rosenblum}
\author[1]{Iván Díaz\thanks{corresponding author:
    ild2005@med.cornell.edu}}
\date{\today}

\affil[1]{\small Division of Biostatistics, Department of Population
  Health Sciences, Weill Cornell Medicine.}
\affil[2]{\small Department of Biostatistics, Johns Hopkins Bloomberg
  School of Public Health.}
\newcommand{\one}{\mathds{1}}

\newcommand{\dd}{\mathrm{d}}
\newcommand{\Stmle}{\tilde S_{\mathrm{TMLE}}}
\newcommand{\Sietmle}{\tilde S_{\mathrm{IE-TMLE}}}
\renewcommand{\P}{\mathsf{P}}
\renewcommand{\lor}{\mathsf{LOR}}
\newcommand{\mw}{\mathsf{MW}}
\newcommand{\rmst}{\mathsf{RMST}}
\newcommand{\rd}{\mathsf{RD}}

\newcommand{\re}{\mathsf{RE}}

\newcommand\indep{\protect\mathpalette{\protect\independenT}{\perp}}
\def\independenT#1#2{\mathrel{\rlap{$#1#2$}\mkern2mu{#1#2}}}

\AtEndDocument{\refstepcounter{theorem}\label{finalthm}}
\AtEndDocument{\refstepcounter{equation}\label{finaleq}}

\title{\titlepaper}

\begin{document}

\maketitle

\begin{abstract}
  The rapid finding of effective therapeutics requires
  the efficient use of available resources in clinical trials. The use
  of covariate adjustment can yield statistical estimates with improved precision, resulting in a reduction in the number of participants
  required to draw futility or efficacy conclusions. We focus on time-to-event and ordinal outcomes. When more than a few baseline covariates are available, a
  key question for covariate adjustment in randomized studies is how
  to fit a model relating the outcome and the baseline covariates to maximize precision.
  We present a novel theoretical result establishing conditions for asymptotic normality
  of a variety of covariate-adjusted estimators that rely on machine learning (e.g., $\ell_1$-regularization,
  Random Forests, XGBoost, and Multivariate Adaptive Regression
  Splines), under the assumption that outcome data is missing completely at random. We further present a consistent estimator of the asymptotic variance. Importantly, the conditions do not require the machine learning methods to converge to the true outcome distribution conditional on baseline variables, as long as they converge to some (possibly incorrect) limit.  
  We conducted a simulation study to evaluate the performance
  of the aforementioned prediction methods  in COVID-19 trials
  using longitudinal data from over 1,500 patients hospitalized with
  COVID-19 at Weill Cornell Medicine New York Presbyterian
  Hospital. We found that using $\ell_1$-regularization led to estimators and corresponding hypothesis tests that control type 1 error and are more
  precise than an unadjusted estimator across all sample sizes
  tested. We also show that when covariates are not prognostic of the
  outcome, $\ell_1$-regularization remains as precise as the
  unadjusted estimator, even at small sample sizes ($n=100$). We give an R package \texttt{adjrct}  that  performs  model-robust covariate adjustment for ordinal and
time-to-event outcomes. 
\end{abstract}

\section{Introduction}

Coronavirus disease 2019 (COVID-19) has affected more than 125 million
people and caused more than 2.7 million deaths worldwide
\citep{whocovid}.  Governments and scientists around the globe have
deployed an enormous amount of resources to combat the pandemic with
remarkable success, such as the development in record time of highly
effective vaccines to prevent disease \citep[e.g.,][]{polack2020safety,
  baden2021efficacy}. Global and local organizations are
launching large-scale collaborations to collect robust scientific data
to test potential COVID-19 treatments, including the testing of drugs
re-purposed from other diseases as well as new compounds
\citep{kupferschmidt2020race}. For example, the World Health Organization launched the SOLIDARITY trial, enrolling almost 12,000
patients in 500 hospital sites in over 30 countries
\citep{who2021repurposed}. Other large initiatives include the
RECOVERY trial \citep{group2020dexamethasone} and the ACTIV initiative
\citep{collins2020accelerating}. To date, there are approximately 2,400 randomized
trials for the treatment of COVID-19 registered in
\url{clinicaltrials.gov}.

The rapid finding of effective therapeutics for COVID-19 requires the
efficient use of available resources. One area where such efficiency
is achievable at little cost is in the statistical design and analysis
of the clinical trials. Specifically, a statistical technique known
as \emph{covariate adjustment} may yield  estimates with
 increased precision (compared to unadjusted estimators), and may result in a reduction of the time, number of
participants, and resources required to draw futility or efficacy
conclusions. This results in faster trial designs, which may help accelerate the
delivery of effective treatments to patients who need them (and may help rule out ineffective treatments faster).

Covariate adjustment refers to pre-planned analysis methods that use  data on patient
baseline characteristics to correct for chance imbalances across study arms, thereby yielding more precise treatment effect estimates. The ICH E9 Guidance on Statistical Methods for Analyzing
Clinical Trials \citep{ICH9} states that ``Pretrial deliberations
should identify those covariates and factors expected to have an
important influence on the primary variable(s), and should consider
how to account for these in the analysis to improve precision and to
compensate for any lack of balance between treatment groups.'' Even
though its benefits can be substantial, covariate adjustment is underutilized; only 24\%-34\% of trials  use covariate adjustment
\citep{kahan2014risks}.  

We focus on estimation of marginal treatment effects, defined as a contrast between study arms in the marginal distribution of the outcome. Many approaches for estimation of marginal treatment effects using covariate adjustment in randomized trials invoke a model relating the outcome and the baseline covariates within strata of treatment. Recent decades have seen a surge in research on the development of
\emph{model-robust} methods for estimating marginal effects that
remain consistent even if this outcome regression model is arbitrarily misspecified
\citep[e.g., ][]{yang2001efficiency,tsiatis2008covariate,
  zhang2008improving,moore2009covariate,
  austin2010substantial,zhang2010increasing,benkeser2020improving}. We focus on a study of the model-robust covariate
adjusted estimators for time-to-event 
 and ordinal outcomes developed by \cite{moore2009covariate}, \cite{diaz2019improved}, and \cite{diaz2016enhanced}. 
  
All potential adjustment covariates  must be
pre-specified in the statistical analysis plan. 
  At the end of the trial, a prespecified prediction algorithm  (e.g., random forests, or using regularization for variable selection) will be run and its output  used to construct a model-robust, covariate adjusted  estimator of the marginal treatment effect for the trial's primary efficacy analysis.
 We aim to address the question of how to do this in a model-robust way that guarantees consistency and asymptotic normality, under some weaker regularity conditions than related work (described below). We also aim to demonstrate  the potential value added by covariate adjustment combined with machine learning, through a simulation study based on COVID-19 data.
 
As a standard regression method for high-dimensional data, $\ell_1$-regularization has been studied by several authors in the context of covariate selection for randomized studies. %For example, \cite{bloniarz2016lasso, tian2012covariate} and \cite{wager2016high} present  conditions that guarantee that an estimator using $\ell_1$-regularization is as or more efficient than the unadjusted estimator; they also give conditions for asymptotic normality of the estimator and  methods for estimating the asymptotic variance. 
For example, \cite{wager2016high} present estimators that are asymptotically normal under strong assumptions that include linearity of the outcome-covariate relationship. \cite{bloniarz2016lasso} present estimators under a randomization inference framework, and show asymptotic normality of the estimators under assumptions similar to the assumptions made in this paper. Both of these papers present results only for continuous outcomes. The method of \cite{tian2012covariate} is general and can be applied to continuous, ordinal, binary, and time-to-event data, and its asymptotic properties are similar to the properties of the methods we discuss for the case of $\ell_1$-regularization, under similar assumptions.
 
More related to our general approach, \cite{wager2016high} also
present a cross-validation procedure that can be used with arbitrary
non-parametric prediction methods (e.g.,
$\ell_1$-regularization, random forests, etc.) in the estimation
procedure. Their proposal amounts to computation of a cross-fitted
augmented inverse probability weighted estimator
\citep{cfVictor}. Their asymptotic normality results, unlike ours, require that that their predictor of the outcome given baseline variables converges to the true regression function.  \citet{WuLOOP2018} proposed a ``leave-one-out-potential outcomes'' estimator where automatic prediction can also be performed using any  regression procedure such as linear regression or random forests, and they propose a conservative variance estimator.  It is unclear as of yet whether Wald-type confidence intervals based on the normal distribution are appropriate for this estimator.  As in the above related work that compares the precision of covariate adjusted estimators to the unadjusted estimator, we assume that outcomes are missing completely at random (since otherwise the  unadjusted estimator is generally inconsistent). 

%We next discuss our contribution to the contribute to the above literature on the theoretical study of general prediction models for covariate adjustment randomized trials. Specifically, 

In Section~\ref{sec:asymp}, we present our main theorem. It shows that any of a large class of prediction algorithms 
(e.g., $\ell_1$-regularization,
  Random Forests, XGBoost, and Multivariate Adaptive Regression
  Splines) 
can be combined with   the covariate adjusted estimator of \cite{moore2009increasing} to produce a consistent,  asymptotically normal estimator of the marginal treatment effect, under regularity conditions. These conditions    do not require consistent estimation of the outcome regression function (as in  key related work described above); instead, our theorem requires the  weaker condition of 
convergence to some (possibly incorrect) limit. We also give a consistent, easy to compute   variance estimator. This has important practical implications because it allows the use machine learning coupled with Wald-type confidence intervals and hypothesis tests, under the conditions of the theorem.  The above estimator can be used with ordinal or time-to-event outcomes.

We next conduct a simulation study to evaluate the performance
of the aforementioned machine learning algorithms for covariate adjustment in the context of COVID-19 trials. We simulate two-arm trials comparing a hypothetical COVID-19
treatment to standard of care. The simulated data distributions are
generated from longitudinal data on approximately 1,500 patients
hospitalized at Weill Cornell Medicine New York Presbyterian Hospital
prior to 15 May 2020. We present results for two types of endpoints:
time-to-event (e.g., time to intubation or death) and ordinal
\citep[e.g., WHO scale, see][]{marshall2020working} outcomes. For
survival outcomes, we present results for two different estimands (i.e., targets of inference): the
survival probability at any given time and the restricted mean
survival time. For ordinal outcomes we present results for the average
log-odds ratio, and for the Mann-Whitney estimand, interpreted as the
probability that a randomly chosen treated patient has a better
outcome than a randomly chosen control patient (with ties broken at random). 

\cite{benkeser2020improving} used simulations based on the above data source to illustrate the efficiency gains achievable by covariate adjustment with parametric models including a small number of adjustment variables (and not using machine learning to improve efficiency). In this paper we evaluate the
performance of four machine learning algorithms
($\ell_1$-regularization, Random Forests, XGBoost, and Multivariate
Adaptive Regression Splines) in several sample sizes, and compare them
in terms of their bias, mean squared error, and type-1 and type-2
errors, to unadjusted estimators and to fully adjusted main terms logistic regression with all available variables included. Furthermore, we
introduce a new R package \texttt{adjrct} \citep{adjrct} that can be
used to perform model-robust covariate adjustment for ordinal and
time-to-event outcomes, and provide R code that can be used to
replicate our simulation analyses with other data sources.

\section{Estimands}\label{sec:estimands}

In what follows, we focus on estimating \textit{intention-to-treat} effects and refer to study arm assignment simply as \textit{treatment}.  We focus on estimation of marginal treatment effects, defined as a contrast between study arms in the marginal distribution of the outcome.  We
further assume that we have data on $n$ trial participants,
represented by $n$ independent and identically distributed copies of
data $O_i:i=1,\ldots,n$. We assume $O_i$ is distributed as $\P$, where
we make no assumptions about the functional form of $\P$ except that 
 treatment is  independent of baseline covariates (by randomization). We denote a generic draw from the distribution $\P$ by $O$. We use the  terms ``baseline covariate'' and ``baseline variable'' interchangeably to indicate a measurement made before randomization.

We are interested in making inferences about a feature of the distribution
$\P$.  We use the word \emph{estimand} to refer to such a feature. We describe example estimands, which include those used in our simulations studies, below. 
%In contrast to regression adjustment, this framework is aligned with the Addendum to the ICH E9 Statistical Principles for Clinical Trials \citep{ICH9addendum}, which states that: ``avoiding or oversimplifying the process of discussing and constructing an estimand risks misalignment between trial objectives, trial design, data collection, and method of analysis. Although an inability to derive a reliable estimate might preclude certain choices of strategy, it is important to proceed sequentially from the trial objective and an understanding of the clinical question of interest, and not for the choice of data collection and method of analysis to determine the estimand.''

\subsection{Ordinal Outcomes}
For ordinal outcomes, assume the observed data is $O=(W,A,Y)$, where
$W$ is a vector of baseline covariates, $A$ is the treatment arm, and
$Y$ is an ordinal variable that can take values in $\{1,\ldots,K\}$. 
Let $F(k, a)=\P(Y\leq k\mid  A=a)$ denote the cumulative
distribution function for patients in arm $A=a$, and let
$f(k, a) = F(k, a)-F(k-1, a)$ denote the corresponding
probability mass function. For notational convenience we will
sometimes use the ``survival'' function instead: $S(k, a)=1-F(k, a)$. The average log-odds ratio is then equal to
\[\lor = \frac{1}{K-1}\sum_{k=1}^{K-1}\log\left[\frac{F(k,
    1)/\{1-F(k, 1)\}}{F(k,
    0)/\{1-F(k, 0)\}}\right],\]
and the Mann-Whitney estimand is equal to
\[\mw=\sum_{k=1}^K\left\{F(k-1, 0) + \frac{1}{2}f(k, 0)\right\}f(k, 1).\]
The Mann-Whitney estimand can be interpreted as the probability that a
randomly drawn patient from the treated arm has a better outcome than
a randomly drawn patient from the control arm, with ties broken at
random \citep{ahmad1996class}. The average log-odds ratio is more
difficult to interpret and we discourage its use, but we include it in
our comparisons because it is a non-parametric extension of the parameter $\beta$ estimated by the
commonly used proportional odds model
$\logit\{F(k, a)\}=\alpha_k + \beta a$ \citep{diaz2016enhanced}.

\subsection{Time to Event Outcomes}
For time to event outcomes, we assume the observed data is
$O=(W,A,\Delta = \one\{Y\leq C\}, \widetilde Y = \min(C,Y))$, where $C$ is
a right-censoring time denoting the time that a patient is last seen, and 
$\one\{E\}$ is the  indicator variable taking the value 1 on the event
$E$ and 0 otherwise. We further assume that events are
observed at discrete time points $\{1,\ldots,K\}$ (e.g., days) as is
typical in clinical trials. The difference in restricted mean survival
time is given by
\[\rmst =\sum_{k=1}^{K-1} \{S(k, 1) - S(k, 0)\},\]
and can be interpreted as a contrast comparing the expected survival
time within the first $K$ time units for the treated arm minus the
control arm \citep{chen2001causal,royston2011use}. The risk difference
at a user-given time point $k$ is defined as
\[\rd = S(k, 1) - S(k, 0),\]
and is interpreted as the difference in survival probability for a
patient in the treated arm minus the control arm.  We note that the
$\mw$ and $\rd$ parameters may be meaningful for both ordinal and
time-to-event outcomes.

\section{Estimators}

For the sake of generality, in what follows we use a common data structure
$O=(W,A,\Delta=\one\{Y\leq C\},\widetilde Y)$ for both ordinal and survival outcomes,
where for ordinal outcomes $C=K$ if the outcome is observed and $C=0$ if it is missing. 

Many approaches for estimation of marginal treatment effects using covariate adjustment in randomized trials invoke a model relating the outcome and the baseline covariates within strata of treatment. It is important that the consistency and interpretability of the treatment effect estimates do not rely on the ability to correctly posit such a model. Specifically, in a recent draft guidance  \citep{FDAguidance}, the FDA states: ``Sponsors can perform covariate adjusted estimation and inference for an unconditional treatment effect ... in the primary analysis of data from a randomized trial. The method used should provide valid inference under approximately the same minimal statistical assumptions that would be needed for unadjusted estimation in a randomized trial.'' The assumption of a correctly specified model is not typically part of the assumptions needed for an unadjusted analysis, and should therefore be avoided when possible. 

All
estimands described in this paper can be computed from the cumulative
distribution functions (CDF) $F(\cdot, a)$ for $a\in\{0,1\}$, which
can be estimated using the empirical cumulative distribution function
(ECDF) or the Kaplan-Meier estimator. Model-robust, covariate
adjusted estimators have been developed for the CDF, including, e.g., 
\cite{chen2001causal,
  rubin2008empirical,moore2009increasing,Stitelman2011,lu2011semiparametric,brooks2013,Zhang2014,Parast2014,benkeser2018improved,diaz2019statistical}. 
 
  We focus on the model-robust, covariate adjusted estimators of \cite{moore2009increasing},
\cite{diaz2016enhanced}, and \cite{diaz2019improved}. 
These estimators
have at least two advantages compared to unadjusted estimators based
on the ECDF or the Kaplan-Meier estimator. First, with time-to-event
outcomes, the adjusted estimator can achieve consistency under an
assumption of censoring being independent of the outcome given study
arm and baseline covariates ($C\indep Y \,|\, A,W$), rather than
the assumption of censoring in each arm being independent of the
outcome marginally ($C\indep Y \,|\, A$) required by unadjusted
estimators. The former assumption is arguably more likely to hold in
typical situations where patients are lost to follow-up due to reasons
correlated with their baseline variables. Second, in large samples and
under regularity conditions, the adjusted estimators of \cite{diaz2016enhanced} and \cite{diaz2019improved} can be at least as
precise as the unadjusted estimator (this requires that missingness/censoring is completely at random, i.e., that in each arm $a \in \{0,1\}$, $C\indep (Y,W)|A=a$), under additional assumptions. 

Additionally, under
regularity conditions, the three aforementioned adjusted estimators are asymptotically normal. This allows the construction of Wald-type confidence
intervals and corresponding tests of the null hypothesis of no
treatment effect.

\subsection{Prediction algorithms}\label{sec:varsel}

While we make no
assumption on the functional form of the distribution $\P$ (except that treatment is independent of baseline variables by randomization),
implementation of our estimators requires a \textit{working model} for the following conditional probability
$$m(k,a,W)=\P(\widetilde Y = k, \Delta = 1\mid \widetilde Y \geq k, A=a, W).$$ 
In time-to-event analysis, this
probability is known as the conditional hazard. The expression
\emph{working model} here means that we do not assume that the
model represents the true relationship between the outcome and the
treatment/covariates. 
Fitting a working model for $m$ is equivalent to training a prediction model for $m$ (specifically, a prediction model for the probability of $\widetilde Y = k, \Delta = 1$ given $\widetilde Y \geq k, A=a, W$), and we sometimes refer to the model fit as a predictor. 

%, but we use the model to capture the association between treatment and outcome within each treatment arm. 
In our
simulation studies, we will use the following working models, fitted in a dataset where each participant contributes a row of data corresponding to each time $k=1$ through $k=\widetilde{Y}$:
\begin{itemize}
\item The following pooled main terms logistic regression (LR)
  $\logit\{m_\beta(k,a,W)\} =\beta_{a,0,k} +\beta_{a,1}^\top W$ estimated with
  maximum likelihood estimation. Note that this model has (i) separate parameters for each study arm, and (ii) in each arm, intercepts for each possible outcome level $k$.
\item The above model fitted with an $\ell_1$ penalty on the parameter
  $\beta_{a,1}$ \citep[$\ell_1$-LR,][]{tibLASSO,park2007l1}.
\item A random forest classification model  \citep[RF,][]{breiman2001random}.
\item An extreme gradient boosting tree ensemble 
  \citep[XGBoost,][]{friedman2001greedy}.
\item Multivariate adaptive regression splines \citep[MARS,][]{friedman1991multivariate}.
\end{itemize}

For RF, XGBoost, and MARS, the algorithms are trained in the whole sample
  $\{1,\ldots,n\}$. For these algorithms, we also assessed the performance of
cross-fitted versions of the estimators. Cross-fitting is sometimes
necessary to guarantee that the regularity assumptions required for
asymptotic normality of the estimators hold when using
data-adaptive regression methods
\citep{klaassen1987consistent,zheng2011cross,cfVictor}, and is
performed as follows. Let ${\cal V}_1, \ldots, {\cal V}_J$ denote a
random partition of the index set $\{1, \ldots, n\}$ into $J$
prediction sets of approximately the same size. That is,
${\cal V}_j\subset \{1, \ldots, n\}$;
$\bigcup_{j=1}^J {\cal V}_j = \{1, \ldots, n\}$; and
${\cal V}_j\cap {\cal V}_{j'} = \emptyset$. In addition, for each $j$,
the associated training sample is given by
${\cal T}_j = \{1, \ldots, n\} \setminus {\cal V}_j$. Let $\widehat m_j$
denote the prediction algorithm trained in $\mathcal T_j$. Letting $j(i)$
denote the index of the prediction set which contains observation $i$,
cross-fitting entails using only observations in $\mathcal T_{j(i)}$
for fitting models when making predictions about observation $i$. That
is, the outcome predictions for each subject $i$ are given by $\widehat
m_{j(i)}(u,a,W_i)$. We let $\widehat\eta_{j(i)}=(\widehat m_{j(i)}, \widehat \pi_A, \widehat \pi_C)$ for cross-fitted estimators and $\widehat\eta_{j(i)}=(\widehat m, \widehat \pi_A, \widehat \pi_C)$ for non-cross-fitted ones. 
RF, XGBoost, and  MARS were fit using the \textit{ranger} \citep{rangerRpac}, \textit{xgboost} \citep{xgboostRpac}, and \textit{earth} \citep{earthRpac} R packages, respectively. Hyperparameter tuning was performed using cross-validation with the \textit{origami} \citep{origamiRpac} R package.

\subsection{Targeted minimum loss based estimation (TMLE)} \label{sec:TMLE}

Our simulation studies use the TMLE procedure presented in \cite{diaz2019improved}. We will refer to that estimator as TMLE with improved efficiency, or IE-TMLE. We will first present the TMLE of \citep{moore2009increasing}, which constitutes the basis for the construction of the IE-TMLE. 
%We now provide a brief overview of the adjusted estimators of $F(k, a)$ are computed for each $(k,a)$ pair in our simulation studies. which are the estimators of . We will refer to those estimators as targeted EF-TMLEWe first define the estimator \citep{moore2009increasing}, which is based on the targeted minimum loss-based framework \citep{vanderLaanRubin2006,van2011targeted,van2018targeted}, and we then define the modification of that estimator that

In the supplementary materials we present some of the efficiency theory underlying the construction of the TMLE. Briefly, TMLE is a framework to construct estimators $\widehat\eta_{j(i)}$ that solve the efficient influence function estimating equation $n^{-1}\sum_{i=1}^n D_{\widehat\eta_{j(i)}}(O_i)=0$, where $D_{\eta}(O)$ is the efficient influence function for $S(k,a)$ in the non-parametric model that only assumes treatment $A$ is independent of baseline variables $W$ (which holds by design), defined in the supplementary materials. TMLE enjoys desirable properties such as local efficiency, outcome model robustness under censoring completely at random, and asymptotic normality, under regularity assumptions.

\textbf{TMLE estimator definition:} Given a predictor $\widehat m$ constructed as in the previous subsection and any $k,a$, 
the corresponding TMLE estimation procedure for $F(k,a)$ can be summarized in the next steps:
\begin{enumerate}
\item Create a long-form dataset where each participant $i$ contributes  the following row of data corresponding to each time $u=0$ through $k$: 
$$\left(u,W_i,A_i, 1\{\widetilde{Y} \geq u\},1\{\widetilde{Y} = u, \Delta=0\}, 1\{\widetilde{Y}=u,\Delta=1\}\right),$$
where $1\{X\}$ is the indicator variable taking value 1 if $X$ is true and $0$ otherwise.
\item For each individual $i$, obtain a prediction $\widehat m(u,a,W_i)$ for
  each pair in the set $\{(u,a):  a = 0, 1; u = 0, \dots, k\}$.
\item Fit a model $\pi_A(a,W)$ for the probability $\P(A=a\mid W)$. Note
  that, in randomized trials, this model may be correctly specified by
  a logistic regression $\logit \pi_A(1,W) = \alpha_0 + \alpha_1^\top
  W$. Let $\widehat \pi_A(a,W_i)$ denote the prediction of the model for
  individual $i$.
\item Fit a model $\pi_C(u,a,W)$ for the censoring probabilities
  $\P(\widetilde Y = u, \Delta = 0\mid \widetilde Y \geq u, A=a, W)$. For
  time-to-event outcomes, this is a model for the censoring
  probabilities. For ordinal outcomes, the only possibilities are that $C=0$ (outcome missing) or $C=K$ (outcome observed); in this case we only fit the aforementioned model at $u=0$ and we set  $\pi_C(u,a,W)=0$ for each $u>0$. For either outcome type, if there is no censoring (i.e., if $P(\Delta=1)=1$), then we set $\pi_C(u,a,W)=0$ for all $u$. Let $\widehat \pi_C(u,a,W_i)$ denote the prediction of this
  model for individual $i$, i.e., using the baseline variable  values from individual $i$. 
\item\label{step:iterate} For each individual $i$ and each $u \leq k$, compute a ``clever''
  covariate $H_{Y,k,u}$ as a function of $\widehat m$, 
  $\widehat \pi_A$, and $\widehat \pi_C$ as detailed in the supplementary materials. The outcome model fit $\widehat m$ is then updated
  by fitting the following logistic regression ``tilting" model with single parameter $\epsilon$ and offset based on $\widehat m$:
  \begin{align*}
           \P(\widetilde Y = u, \Delta = 1\mid \widetilde Y \geq u, A=a, W) = 
            \logit^{-1}\left\{\logit \widehat m(u,a,W) + \varepsilon H_{Y,k,u}\right\}.
            \end{align*}
%   We denote the right side of the above display by     $m_\varepsilon(u,a,W)$. 
   This can be done using standard statistical software for fitting a logistic regression of the indicator variable 
      $1\{\widetilde Y = u, \Delta = 1\}$ on the variable $H_{Y,k,u}$ using offset $\logit \widehat m(u,a,W)$ among observations with $\widetilde Y \geq u$ and $A=a$ in the long-form dataset from step 1.
   The above model fitting process is iterated 
   where at the beginning of each iteration we replace $\widehat m$ in the above display and in the definition of $H_{Y,k,u}$  by the updated model fit.
   We denote the maximum number of iterations that we allow by $i_{\max}$.
  % \begin{align*}
  %   \logit m_\varepsilon(u,a,W) &= \logit \widehat m(u,a,W) + \varepsilon
  %                              H_{Y,u}\\
  %   , $H_{A,i}$ and $H_{C,i}$
  %   \logit \pi_{\gamma,A}(1,W) &= \logit \widehat \pi_A(1,W) + \gamma H_A\\
  %   \logit \pi_{\alpha,C}(u,a,W) &= \logit \widehat \pi_C(u,a,W) + \upsilon H_{C,u}
  % \end{align*}
\item Let $\widetilde m(u,a,W_i)$  denote the estimate of $m(u,a,W_i)$ for individual $i$ at the final iteration of the previous step. Note that this estimator is specific to the value $k$ under consideration.
\item Compute the  estimate of $S(k,a)=1-F(k,a)$ as the following standardized estimator
  \begin{equation}
      \Stmle(k, a) = \frac{1}{n}\sum_{i=1}^n \prod_{u=1}^k\{1-\widetilde m(u,a,W_i)\},\label{def:tmle}
  \end{equation}
  and let the estimator of 
  $F(k,a)$ be $1-\Stmle(k, a)$.
\end{enumerate}
This estimator was originally proposed by
\cite{moore2009increasing}. The role of the clever covariate $H_{Y,k,u}$
is to endow the resulting estimator $\widetilde S(k,a)$ with 
properties such as model-robustness  compared
to unadjusted estimators. In particular, it can be shown that this
estimator is efficient when the working model for $m$ is correctly specified. The specific
form of the covariate $H_{Y,k,u}$ is given in the supplementary
materials. 
Throughout, the notation $\widehat m$ is used to represent the predictor constructed as in Section~\ref{sec:varsel} and which is an input to the above TMLE algorithm, while $\widetilde m$ denotes the updated version of  this predictor that is output by the above TMLE algorithm at step 6.

\textbf{IE-TMLE estimator definition:} 
In Section~\ref{sec:methods} we will compare several machine learning
procedures for estimating $m$ in finite samples. The estimators used in the simulation study are the IE-TMLE of
\cite{diaz2019improved}, where in addition to updating the
initial estimator for the outcome regression $m$, we also update the
estimators of the treatment and censoring mechanisms. Specifically, we
replace step \ref{step:iterate} of the above procedure with the following:
\begin{enumerate}
  \hypersetup{linkcolor=black}
\item[\ref{step:iterate}.] For each individual $i$ construct ``clever''
  covariates $H_{Y,k,u}$, $H_A$, and $H_{C,k,u}$ (defined in the supplementary materials) as a function of
  $\widehat m$, $\widehat \pi_A$, and $\widehat
  \pi_C$. For each $k=1,\ldots,K$, the model fits are then
  iteratively updated using logistic regression ``tilting" \hypersetup{linkcolor=red}
  models:
  \begin{align*}
    \logit m_\varepsilon(u,a,W) &= \logit \widehat m(u,a,W) + \varepsilon
                               H_{Y,k,u}\\
    \logit \pi_{\gamma,A}(1,W) &= \logit \widehat \pi_A(1,W) + \gamma H_A\\
    \logit \pi_{\upsilon,C}(u,a,W) &= \logit \widehat \pi_C(u,a,W) + \upsilon H_{C,k,u}
  \end{align*}
  where the iteration is necessary because $H_{Y,k,u}$, $H_{A}$, and $H_{C,k,u}$ are functions of
  $\widehat m$, $\widehat \pi_A$, and $\widehat \pi_C$ that must be updated at each step. As before, for ordinal outcomes we only fit the aforementioned model at $u=0$ and we set  $\pi_C(u,a,W)=0$ for each $u>0$.
\end{enumerate}
We use $\Sietmle$ to denote this estimator. The updating step above combines ideas from \cite{moore2009increasing},
\cite{Gruber2012t}, and \cite{Rotnitzky2012} to produce an estimator
with the following properties:
\begin{enumerate}[(i)]
\item Consistency and at least
  as precise as the Kaplan-Meier and inverse probability weighted
  estimators;
\item Consistency under violations of independent censoring (unlike the
  Kaplan-Meier estimator) when either the censoring or survival
  distributions, conditional on covariates, are estimated
  consistently and censoring is such that $C\indep Y\mid W, A$; and
\item Nonparametric efficiency when both of these distributions are
  consistently estimated at rate $n^{1/4}$.
\end{enumerate}
Please see \cite{diaz2019improved} for more details on these estimators, which are implemented in the R package \texttt{adjrct}
\citep{adjrct}.

Next, we present a result (Theorem~\ref{theo:normal}) stating asymptotic normality of $\Stmle$ using machine learning for prediction that avoids some limitations of existing methods, and present a consistent estimator of its variance. In Section~\ref{sec:methods} we present simulation results where we evaluate the performance of  $\Sietmle$ for covariate adjustment in COVID-19 trials for hospitalized patients. We favor $\Sietmle$ in our numerical studies because, unlike $\Stmle$, it satisfies property (i) above. The simulation uses Wald-type hypothesis tests based on the asymptotic approximation of Theorem~\ref{theo:normal}, where we note that the variance estimator in the theorem is consistent for $\Stmle$ but it is conservative for $\Sietmle$ \citep{moore2009increasing}. 

\subsection{Asymptotically correct confidence intervals and hypothesis tests for  TMLE combined with machine learning}\label{sec:asymp}

Most available methods to construct confidence intervals and hypothesis tests in the statistics literature are based on the sampling distribution of the estimator. While using the exact finite-sample distribution would be ideal for this task, such distributions are notoriously difficult to derive for our problem in the absence of strong and unrealistic assumptions (such as linear models with Gaussian noise). Thus, here we focus on methods that rely on approximating the finite-sample distribution using asymptotic results as $n$ goes to infinity. 

In order to discuss existing methods, it will be useful to introduce and compare the following assumptions:

\begin{assumption}\label{ass:randomcensor}
  Censoring is completely at random, i.e., $C\indep
  (Y,W)\mid A=a$ for each treatment arm $a$.
\end{assumption}

\begin{assumption}\label{ass:consistentm}
  Let $||f||^2$ denote the $L_2(\P)$ norm $\int
  f^2(o)\dd\P(o)$, for $O=(W,A,\Delta=\one\{Y\leq C\},\widetilde Y)$. 
  We abbreviate $m(k,a,W)$ and $\widehat{m}(k,a,W)$ by $m$ and $\widehat{m}$, respectively. 
  Assume the estimator $\widehat{m}$ is consistent in the sense that $||\widehat{m} - m||=o_P(1)$ for all $k\in\{1,\ldots,K\}$ and $a\in\{0,1\}$.
  We also assume that there exists a $\delta>0$ such that $\delta < m < 1-\delta$ with probability 1.
\end{assumption}

\begin{assumption}\label{ass:limitm}
  Assume the estimator $\widehat{m}$ converges to a possibly misspecified limit $m_1$ in the sense that $||\widehat{m} - m_1||=o_P(1)$ for all $k\in\{1,\ldots,K\}$ and $a\in\{0,1\}$, where we emphasize that $m_1$ can be different from the true regression function $m$. We also assume that there exists a $\delta>0$ such that $\delta < m_1 < 1-\delta$ with probability 1. 

\end{assumption}
For estimators $\widehat{m}$ of $m$ that use cross-fitting, the function $\widehat{m}$ consists of $J$ maps (one for each training set) from the sample space of $O$ to the  interval $[0,1]$. In this case, by convention we define $||\widehat{m} - m||$ in \ref{ass:consistentm}  as the average across the $J$ maps of the 
$L_2(\P)$ norm of each such map minus $m$. Convergence of  $||\widehat{m} - m||$ to $0$ in probability is then equivalent to the same convergence where $\widehat{m}$ is replaced by the corresponding map before cross-fitting is applied. The same convention is used in \ref{ass:limitm}.

There are at least two  results on asymptotic normality for $\Stmle$  relevant to the problem we are studying. The first result is a general theorem for TMLE \cite[see Appendix A.1 of][]{van2011targeted}, stating that the estimator  is asymptotically normal and efficient under regularity assumptions which include \ref{ass:consistentm}. Among other important implications, this asymptotic normality implies that the variance of the estimators can be consistently estimated by the empirical variance of the efficient influence function. This means that asymptotically correct confidence intervals and hypothesis tests can be constructed using a Wald-type procedure. As stated above, it is often undesirable to assume \ref{ass:consistentm} in the setting of a randomized trial, as it is a much stronger assumption than what would be required for an unadjusted estimator. 

The second result of relevance to this paper establishes asymptotic normality of $\widetilde S(k,a)$  under assumptions that include \ref{ass:limitm} \citep{moore2009covariate}. The asymptotic variance derived by these authors depends on the true outcome regression function $m$, and is thus difficult to estimate. As a solution, the authors propose to use a conservative estimate of the variance whose computation does not rely on the true regression function $m$. While this conservative method yields correct type 1 error control, its use is not guaranteed to fully covert precision  gains from covariate adjustment into power gains. 

We note that the above asymptotic normality results from related works rely on the additional condition that the estimator $\widehat m$ lies in a Donsker class. This assumption may be violated by some of the data-adaptive regression techniques that we consider. Furthermore, we note that resampling methods such as the bootstrap cannot be safely used for variance estimation in this setting. Their correctness is currently unknown when the working model for $m$ is based on data-adaptive regression procedures such as those described in Section \ref{sec:varsel} and used in our simulation studies. 

In what follows, we build on recent literature on estimation of causal effects using machine learning to improve upon the aforementioned asymptotic normality results on two fronts. First, we introduce cross-fitting \citep{klaassen1987consistent, zheng2011cross,cfVictor} to avoid the Donsker condition. Second, and most importantly, we present a novel asymptotic normality result that avoids the above limitations of existing methods regarding strong assumptions (specifically \ref{ass:consistentm}) and conservative variance estimators (that may sacrifice power). 

The following are a set of assumptions about how components of the TMLE are implemented, which we'll use in our theorem below:

\begin{assumption}\label{ass:ga}
  The initial estimator of $\pi_A(1)$ is set to be the empirical mean $n^{-1}\sum_{i=1}^nA_i$.
\end{assumption}
\begin{assumption}\label{ass:gc}
  For time-to-event outcomes, the initial estimator $\widehat \Pi_C(a, u)$ is set to be the Kaplan-Meier
  estimator estimated separately within each treatment arm $a$. For ordinal outcomes, $\widehat \Pi_C(a, 0)$ is the proportion of missing outcomes in treatment arm $a$ and $\widehat \Pi_C(a, u)=0$ for $u>0$.
\end{assumption}
\begin{assumption}\label{ass:m}
  The initial estimator $\widehat m(u,a,W)$ is constructed using one of the following:
  \begin{enumerate}
  \item Any estimator in a parametric working model (i.e., a model that can be indexed by a Euclidean parameter) such as maximum likelihood, $\ell_1$ regularization, etc.\label{case:param}
  \item Any data-adaptive regression method (e.g., random forests,
    MARS, XGBoost, etc.) estimated using cross-fitting as described above.\label{case:cross}
  \end{enumerate}
  \end{assumption}
\begin{assumption} \label{ass:regularity_for_consistency}
  The regularity conditions in Theorem~5.7 of \citep[p.45]{vanderVaart98} hold for the maximum likelihood estimator corresponding to each logistic regression model fit in step (\ref{step:iterate}) of the TMLE algorithm.
%This theorem then implies convergence of the estimated $\epsilon$ from step (\label{step:iterate}) of the TMLE algorithm to a finite limit $\epsilon_1$.
\end{assumption}

\begin{theorem}\label{theo:normal}
Assume \ref{ass:randomcensor} and  \ref{ass:limitm}--\ref{ass:regularity_for_consistency} above. Define the variance estimator
\[\widetilde\sigma^2 = \frac{1}{n}\sum_{i=1}^n
  [D_{\widetilde\eta_{j(i)}}(O_i)]^2.\]Then we have for all $k\in\{1,\ldots,K\}$ and $a\in\{0,1\}$ that 
  \[\sqrt{n}\{\Stmle(k,a) -S(k,a)\}/\widetilde\sigma \rightsquigarrow N(0,1).\]
\end{theorem}

Theorem~\ref{theo:normal} is a novel  result establishing the asymptotic
correctness of Wald-type confidence intervals and hypothesis tests for
the covariate-adjusted estimator $\Stmle(k,a)$ based on machine learning regression procedures
constructed as stated in \ref{ass:m}. For example, the confidence interval $\Stmle(k,a)\pm 1.96\times \widetilde\sigma/\sqrt{n}$ has approximately 95\% coverage at large sample sizes, under the assumptions of the theorem. The theorem
licenses the large sample use of any regression procedure for $m$ when combined  with the TMLE of Section~\ref{sec:TMLE}, as long as the regression procedure is either (i) based on
a parametric model (such as $\ell_1$-regularization) or (ii) based on
cross-fitted data-adaptive regression, and the assumptions of the theorem hold. The theorem states sufficient assumptions under which Wald-type tests from such a procedure will be asymptotically correct. 

Assumption \ref{ass:limitm} states that the predictions given by the regression method used to construct the adjusted estimator converge to some arbitrary function (i.e., not assumed to be equal to the true regression function). This assumption is akin to Condition 3 assumed by \cite{bloniarz2016lasso} in the context of establishing asymptotic normality of a covariate-adjusted estimator based on $\ell_1$-regularization. We note that this is an assumption on the predictions themselves and not on the functional form of the predictors. Therefore, issues like collinearity do not necessarily cause problems. While this assumption can hold for many off-the-shelf machine learning regression methods under assumptions on the data-generating mechanism, general conditions have not been established and the assumption must be checked on a case-by-case basis. %Results of this type can be proved  have already been established for some estimators; for example, \cite{lu2012robustness} presents conditions under which \ref{ass:limitm} holds for the adaptive lasso \citep{zou2006adaptive} under a semi-parametric model, and \cite{athey2019generalized} present results for random forests. 

We note that assumption \ref{ass:randomcensor} is stronger than the assumption $C\indep Y \mid A=a$ required by unadjusted estimators such as the Kaplan-Meier estimator. However, if $W$ is prognostic (meaning that $W\not\indep Y\mid A=a$), then the assumption $C\indep Y \mid A=a$ required by the Kaplan-Meier estimator cannot generally be guaranteed to hold, unless \ref{ass:randomcensor} also holds. Thus, our theorem aligns with the recent FDA  draft guidance on covariate adjustment in the sense that ``it provides valid inference under approximately the same minimal statistical assumptions that would be needed for unadjusted estimation in a randomized trial'' \citep{FDAguidance}. 

The construction of estimators based on \ref{ass:gc} should be avoided if \ref{ass:randomcensor} does not hold. Confidence that \ref{ass:randomcensor} holds is typically warranted in trials where  the only form of right censoring is administrative. When applied to ordinal outcomes, \ref{ass:randomcensor}  is trivially satisfied if there is no missing outcome data. 

Consider the case where censoring is informative such that \ref{ass:randomcensor} does not hold, but censoring at random holds (i.e., $C\indep Y\mid W, A$).  Then consistency of the estimators $\Stmle$ and $\Sietmle$ will typically require that \emph{at least one} of two assumptions hold: (a) that the censoring probabilities $\pi_C(u,a,w)$ are estimated consistently, or that (b) the outcome regression $m(u,a,w)$ is estimated consistently. To maximize the chances of either of these conditions being true, we recommend the use of flexible machine learning for both of these regressions, including model selection and ensembling techniques such as the Super Learner \citep{van2007super}. The conditions for asymptotic normality of  $\Stmle$ and $\Sietmle$ under these circumstances are much stronger than those for Theorem~\ref{theo:normal}, and typically include consistent estimation of \emph{both} $\pi_C(u,a,w)$ and $m(u,a,w)$ at certain rates \citep[e.g., each of them converging at $n^{1/4}$-rate is sufficient, see Appendix A.1 of][]{van2011targeted}.

\section{Simulation methods}\label{sec:methods}

Our data generating distribution is based on a database of over 1,500
patients hospitalized at Weill Cornell Medicine New York Presbyterian
Hospital prior to 15 May 2020. The database includes information on
patients 18 years of age and older with COVID-19 confirmed through
reverse-transcriptase–polymerase-chain-reaction assays.  For a full
description of the clinical characteristics and data collection
methods of the initial cohort sampling, see \citet{goyalNEJM}.  

We evaluate the potential to improve efficiency by adjustment for subsets of the following baseline
variables: age, sex, BMI, smoking status, whether the patient required
supplemental oxygen within three-hours of presenting to the emergency
department, number of comorbidities (diabetes, hypertension, COPD,
CKD, ESRD, asthma, interstitial lung disease, obstructive sleep apnea,
any rheumatological disease, any pulmonary disease, hepatitis or HIV,
renal disease, stroke, cirrhosis, coronary artery disease, active
cancer), number of relevant symptoms, presence of bilateral
infiltrates on chest x-ray, dyspnea, and hypertension. These variables
were chosen because they have been previously identified as risk
factors for severe disease \citep{chinacovid, goyalNEJM,
  Gupta2003498}, and therefore are likely to improve efficiency of
covariate-adjusted effect estimators in randomized trials in hospitalized patients.

Code to reproduce our simulations may be found at \url{https://github.com/nt-williams/covid-RCT-covar}.

\subsection{Data generating mechanisms}

We consider two types of outcomes: a time-to-event outcome defined as
the time from hospitalization to intubation or death, and a six-level
ordinal outcome at 14 days post-hospitalization based on the WHO
Ordinal Scale for Clinical Improvement
\citep{marshall2020working}. The categories are as follows: 0,
discharged from hospital; 1, hospitalized with no oxygen therapy; 2,
hospitalized with oxygen by mask or nasal prong; 3, hospitalized with
non-invasive ventilation or high-flow oxygen; 4, hospitalized with
intubation and mechanical ventilation; 5, dead. For time to event
outcomes, we focus on evaluating the effect of treatment on the
$\rmst$ at 14 days and the $\rd$ at 7 days after hospitalization, and
for ordinal outcomes we evaluate results for both the $\lor$ and the
Mann-Whitney statistic.

We simulate datasets for four scenarios where we consider two effect
sizes (null versus positive) and two baseline variable settings
(prognostic versus not prognostic, where prognostic means marginally associated with the outcome). For each sample size
$n \in \{100, 500, 1500\}$ and for each scenario, we simulated $5000$
datasets as follows. To generate datasets where covariates are
prognostic, we draw $n$ pairs $(W,Y)$ randomly from the original
dataset with replacement. This generates a dataset where the covariate
prognostic strength is as observed in the real dataset. To simulate
datasets where covariates are not prognostic, we first draw outcomes
$Y$ at random with replacement from the original dataset, and then
draw covariates $W$ at random with replacement and independent of the
value $Y$ drawn. 

For each scenario, a hypothetical treatment variable
is assigned randomly for each patient with probability $0.5$ independent of all other variables. This produces a data generating distribution with zero treatment effect. Next, a positive  treatment effect is simulated for time-to-event outcomes by
adding an independent random draw from a $\chi^2$ distribution four
degrees of freedom to each patient's observed survival time in the
treatment arm. This effect size translates to a difference in $\rmst$
of 1.04 and $\rd$ of 0.10, respectively. To simulate outcomes being missing completely at random, $5\%$ of the patients are selected at random to be
censored, and the censoring times are drawn from a uniform distribution
between $1$ and $14$. A positive treatment effect is simulated for
ordinal outcomes by subtracting from each patient's outcome in the treatment arm an independent
random draw from a four-parameter Beta distribution with support
$(0, 5)$ and parameters $(3, 15)$, rounded to the nearest nonnegative
integer. This generates effect sizes for $\lor$ of 0.60 and for $\mw$ of 0.46.

\section{Simulation results}\label{sec:results}

We evaluate several estimators.  First, we evaluate unadjusted
estimators based on substituting the empirical CDF for ordinal outcomes and the
Kaplan-Meier estimator for time-to-event outcomes in the parameter definitions of Section~\ref{sec:estimands}. We then evaluate
adjusted estimator $\Sietmle(k,a)$ where the working models are: 
\begin{itemize}[noitemsep,leftmargin=2cm]
    \item[LR:] a fully adjusted
    estimator using logistic regression including all the variables listed
    in the previous section,
    \item[$\ell_1$-LR:] $\ell_1$ regularization of the previous
logistic regression,
\item[RF:] random forests,
\item[MARS:] multivariate adaptive
regression splines, and 
\item[XGBoost:] extreme gradient boosting tree ensembles.
\end{itemize}
For estimators RF, MARS, and XGBoost, we further evaluated cross-fitted
versions of the working model. For all adjusted estimators the
propensity score $\pi_A$ is estimated with an intercept-only model (\ref{ass:ga}),
and the censoring mechanism $\pi_C$ is estimated using a Kaplan-Meier estimator fitted independently for each treatment arm (\ref{ass:gc}) (or equivalently for ordinal outcomes the proportion of missing outcomes within each treatment arm). 

Confidence intervals and hypothesis tests are
performed using Wald-type statistics, which use an estimate of the standard error. The standard error was estimated based on the asymptotic Gaussian
approximation described in Theorem \ref{theo:normal}. We compare the performance of
the estimators in terms of the probability of type-1 error, power, the absolute bias, the variance, and the
mean squared error. 

We compute the relative efficiency $\re$ of each
estimator compared to the unadjusted estimator as a ratio of the mean
squared errors. This relative efficiency can be interpreted as the
ratio of sample sizes required by the estimators to achieve the same
power at local alternatives, asymptotically \citep{vaart_1998}. Equivalently, one minus the relative efficiency is the relative reduction (due to covariate adjustment) in the required sample size to achieve a desired power, asymptotically; e.g., a relative efficiency of 0.8 is approximately  equivalent  to needing 20\% smaller sample size when using covariate adjustment.

In the presentation of the results, we
append the prefix CF to cross-fitted estimators. For example, CF-RF
will denote cross-fitted random forests. % We use $\ell_1$-LR to
% refer to $\ell_1$-regularized logistic regression, and LR to refer to
% fully adjusted logistic regression, i.e.,  (a) above.

Tables containing the comprehensive results of the simulations are
presented in the supplementary materials. In the remainder of this section we present a summary of the results. First, we note that the use of
random forests without cross-fitting exhibits very poor performance,
failing to appropriately control type-1 error when the effect is null,
and introducing significant bias when the effect is positive. We
observed this poor performance across all simulations. Thus, in what follows
we omit a discussion of this estimator. 

Results for the $\lor$ in Tables~\ref{tab:lor:yy} and \ref{tab:lor:yn} show that covariate adjusted estimators have better performance than the unadjusted estimator at small sample sizes, even when the covariates are not prognostic. In these cases, the unadjusted estimator is unstable with large variance due to near-empty outcome categories in some simulated datasets, which causes division by
near-zero numbers in the unadjusted $\lor$ estimator. Some covariate
adjusted estimators fix this problem by extrapolating model
probabilities to obtain better estimates of the probabilities in the
near-empty cells.

Tables \ref{tab:rmst:yy}-\ref{tab:mw:yy} (in the web supplementary
materials) display the results for the
difference in $\rmst$, $\rd$, $\lor$, and $\mw$ estimands when covariates are
prognostic and there is a positive effect size.  At sample size
$n=1500$ all adjusted estimators yield efficiency gains, with CF-RF
offering the best $\re$ ranging from $0.51$ to $0.67$ compared to an
unadjusted estimator, while appropriately controlling type-1
error. In contrast, the $\re$ of $\ell_1$-LR at
$n=1500$ ranged from $0.79$ to $0.89$. 

At sample size $n=500$,
$\ell_1$-LR, CF-RF, and XGBoost offer comparable efficiency gains,
ranging from $0.29$ to $0.99$. As the sample size decreases to $n=100$
most adjusted estimators yield efficiency losses and the only
estimator that retains efficiency gains is $\ell_1$-LR, with $\re$
from $0.86$ to $0.92$. (An exception is in estimation of the
$\lor$, where the $\re$ of $\ell_1$-LR was $0.1$ due to the issue discussed above.) 

Efficiency gains for
$\ell_1$-LR did not always translate into power gains of a Wald-type
hypothesis test compared to other estimators (e.g. LR at $n=100$),
possibly due to biased variance estimation and/or a poor Gaussian
approximation of the distribution of the test statistic. At small
sample size $n=100$ power was uniformly better for a Wald-type test
based on LR compared to $\ell_1$-LR. At sample size $n=500$ a Wald-type test based on
$\ell_1$-LR seemed to dominate all other algorithms, whereas at
$n=1500$ all algorithms had comparable power very close to one.

Results when the true treatment effect is zero and covariates are prognostic are presented
in Tables \ref{tab:rmst:ny}-\ref{tab:mw:ny}  (in the web supplementary
materials). At sample size $n=1500$,
CF-RF generally provides large efficiency gains with relative
efficiencies ranging from $0.66$ to $0.77$. For comparison,
$\ell_1$-LR has $\re$ ranging from $0.88$ to $0.92$. As the
sample size decreases to $n=500$, $\ell_1$-LR and CF-RF both offer the
most efficiency gains while retaining type-1 error control, with $\re$
ranging from $0.74$ to $0.88$. At small sample sizes $n=100$,
$\ell_1$-LR consistently leverages efficiency gains from covariate
adjustment ($\re$ ranging from $0.73$ to $0.95$) but its type-1 error (ranging
from $0.07$ to $0.09$) is slightly larger than that of the unadjusted estimator. For estimation of $\lor$ and $\mw$, XGBoost has similar results at sample size $n=100$.

Tables \ref{tab:rmst:yn}-\ref{tab:mw:yn} (in the web supplementary
materials) show results for scenarios
where the covariates are not prognostic of the outcome but there is a
positive effect. This case is interesting because it is well known
that adjusted estimators can induce efficiency losses (i.e., $\re >1$) by adding
randomness to the estimator when there is nothing to be gained from
covariate adjustment. We found that $\ell_1$-LR uniformly avoids
efficiency losses associated with adjustment for independent
covariates, with a maximum $\re$ of $1.03$. All other covariate
adjustment methods had larger maximum $\re$. At sample size $n=100$,
the superior efficiency of the $\ell_1$-LR estimator did not always
translate into better power (e.g., compared to LR) due to the use of a
Wald-test which relies on an asymptotic approximation to the distribution of the estimator.

Results when the true treatment effect is zero and covariates are not prognostic are presented in
Tables~\ref{tab:rmst:nn}-\ref{tab:mw:nn}  (in the web supplementary
materials). In this case, $\ell_1$-LR
also avoids efficiency losses across all scenarios, while maintaining
a type-1 error that is comparable to that of the unadjusted estimator.

Lastly, at large sample sizes all cross-fitted estimators along with
logistic regression estimators yield correct type I error,
illustrating the correctness of Wald-type tests proved in
Theorem~\ref{theo:normal}. Our simulation  results also show that
Wald-type hypothesis tests based on data-adaptive machine learning
procedures fail to control type 1 error if the regressions procedures are not
cross-fitted.

\section{Recommendations and future directions}\label{sec:recs}

In our numerical studies we found that $\ell_1$-regularized logistic
regression offers the best trade-off between type-I error control and
efficiency gains across sample sizes, outcome types, and
estimands. We found that this algorithm leverages efficiency gains when efficiency gains are feasible, while protecting the estimators from efficiency losses when efficiency gains are not feasible (e.g., adjusting for covariates with no prognostic power).  A direction of future research is the evaluation of bootstrap estimators for the variance and confidence intervals of covariate-adjusted estimators, especially for cases where the Wald-type methods evaluated in this manuscript did not perform well (e.g., $\ell_1$-LR at $n=100$). 

We also found that logistic regression can result in large efficiency losses for
small sample sizes, with relative efficiencies as large as $1.17$ for
the $\rmst$ estimand, and as large as $7.57$ for the $\mw$
estimand. Covariate adjustment with $\ell_1$-regularized logistic
regression solves this problem, maintaining efficiency when covariates
are not prognostic for the outcome, even at small sample
sizes. However, Wald-type hypothesis tests do not appropriately
translate the efficiency gains of $\ell_1$-regularized logistic
regression into more powerful tests. This requires the development of
tests appropriate for small samples.

We recommend against using the $\lor$ parameter since it is difficult
to interpret and the corresponding estimators (even unadjusted ones)
can be unstable at small sample sizes. Covariate adjustment with
$\ell_1$-LR, CF-MARS, CF-RF, or CF-XGBoost can aid to improve
efficiency in estimation of the $\lor$ parameter over the unadjusted estimator
when there are near-empty cells at small sample sizes. This
improvement in efficiency did not translate into an improvement in
power when using Wald-type hypothesis tests, due to poor small-sample
Gaussian approximations or poor variance estimators.

We discourage the use of non-cross-fitted versions of the machine learning methods evaluated (i.e., RF, XGBoost, MARS) for covariate adjustment. Specifically, we found  in simulations that non-cross-fitted random forests can lead to
overly biased estimators in the case of a positive effect, and to
anti-conservative Wald-type hypothesis tests in the case of a null
treatment effect. We found that cross-fitting the random forests
alleviated this problem and was able to produce small bias and
acceptable type-1 error at all sample sizes. This is supported at large sample sizes by our main theoretical result (Theorem~\ref{theo:normal}) which establishes asymptotic correctness of cross-fitted procedures under regularity conditions. In fact, we found that
random forests with cross-fitting provided the most efficiency gains
at large sample sizes. 

Based on the results of our simulation studies,
we recommend that cross-fitting with data-adaptive estimators such as
random forests and extreme gradient boosting be considered for covariate
selection in trials with large sample sizes ($n=1500$ in our simulations). In large sample sizes, it is also possible to consider  an ensemble approach such as Super Learning \citep{van2007super} that allows one to select the predictor that yields the most efficiency gains. Traditional model selection with statistical learning is focused on the goal of prediction, and an adaptation of those tools to the goal of maximizing efficiency in estimating the marginal treatment effect is the subject of future research.

The conditions for asymptotic normality and consistent variance
estimation of $\Stmle(k,a)$ established in Theorem~\ref{theo:normal}
may be restrictive if censoring is informative. In that case,
consistency of the $\Stmle(k,a)$ and $\Sietmle(k,a)$ estimators
requires that censoring at random holds (i.e., $C\indep Y\mid W, A$),
and that either the outcome regression or censoring mechanism is
consistently estimated. Thus, it is recommended to also estimate the
censoring mechanism with machine learning methods that allow for
flexible regression. Standard asymptotic normality results for the
$\Stmle(k,a)$ and $\Sietmle(k,a)$ require consistent estimation of
both the censoring mechanism and the outcome mechanism at certain
rates (e.g., both estimated at a $n^{1/4}$ rate is sufficient). The
development of estimators that remain asymptotically normal under the
weaker condition that at least one of these regressions is
consistently estimated has been the subject of recent research
\citep[e.g.,][]{diaz2017doubly,benkeser2017doubly,diaz2019statistical}.

% \printbibliography
\bibliographystyle{plainnat}
\bibliography{references}

\end{document}

\maketitle

\appendix

\section{Auxiliary covariates for estimation algorithm}
Denote the survival function for $Y$ at time $k \in \{1,\dots, K\}$
conditioned on study arm $a$ and baseline variables $w$ by
\begin{equation}
  S(k,a,w)=P(Y>k\mid A=a,W=w).\label{S_def}
\end{equation}
Similarly, define the following function of the censoring distribution:
\begin{equation}
  G(k,a,w)=P(C\geq k\mid  A=a,W=w). \label{G_def} \end{equation}
Under the assumption $C\indep
(Y,W)\mid A=a$ for each treatment arm $a$, we have $Y \indep C \mid  A,W$ and therefore
$S(k,a,w)$ and $G(k,a,w)$  have the following product formula representations:
\begin{align}
  S(k,a,w)&=\prod_{u=1}^k \{1-m(u,a,w)\}; \qquad
            \Pi_C(k,a,w)=\prod_{u=0}^{k-1} \{1-\pi_C(u,a,w)\}.
            \label{defS}
\end{align}

At each iteration of the estimation algorithm, the auxiliary
covariates fr $\Stmle$ and $\Sietmle$ are constructed as follows:
\begin{align*}
  H_{Y,k,u} = & -\frac{\one\{A=a\}}{\widehat \pi_A(a,W)\widehat \Pi_C(u,a,W)}\frac{\widehat S(k,a,W)}{\widehat S(u,a,W)}\\
  H_A = &\frac{S(k,a,W)}{\pi_A(a,W)},\\
  H_{C,k,u} = & -\frac{\one\{A=a\}}{\widehat \pi_A(a,W)}\frac{\widehat S(k,a,W)}{\widehat S(u,a,W)}\frac{1}{\widehat \Pi_C(u+1,a,W)},
\end{align*}
where $\widehat \pi_A$, $\widehat S$, and $\widehat \Pi_C$ are the estimates in the
current step of the iteration. 

\section{Efficiency theory}
Before proving Theorem \ref{theo:normal} given in the paper, we
introduce some notation and efficiency theory for estimation of
$S(k,a)$. We will use the notation of \cite{diaz2019improved}.
First, we encode a single participant's data vector
$O=(W,A,\Delta = \one\{Y\leq C\}, \widetilde Y = \min(C,Y))$ using the
following longitudinal data structure:
\begin{equation}
  O=(W, A, R_0, L_1,  R_1, L_2\ldots, R_{K-1}, L_K),\label{O}
\end{equation}
where $R_u = \one\{\widetilde Y = u, \Delta=0\}$ and
$L_u= \one\{\widetilde Y = u, \Delta=1\}$, for $u\in\{0,\ldots,K\}$. The
sequence $R_0, L_1, R_1, L_2\ldots, R_{K-1}, L_K$ in the above display
consists of all $0$'s until the first time that either the event is
observed or censoring occurs, i.e., time $u=\widetilde{Y}$. In the former
case $L_u=1$; otherwise $R_u=1$. For a random variable $X$, we denote
its history through time $u$ as $\bar X_u=(X_0,\ldots,X_u)$. For a
given scalar $x$, the expression $\bar X_u=x$ denotes element-wise
equality. The corresponding vector (\ref{O}) for participant $i$ is
denoted by
$(W_i,A_i,R_{0,i}, L_{1,i}, R_{1,i}, L_{2,i}\ldots,
R_{K-1,i},L_{K,i})$.

% We say the event is observed at time $t$ if $T=t$ and $C\geq t$.
Define the following indicator variables for each $u \geq 1$:
$$I_u=\one\{\bar R_{u-1}=0, \bar L_{u-1}=0\}, \qquad
J_u=\one\{\bar R_{u-1}=0, \bar L_u=0\}.$$ The variable $I_u$ is the
indicator based on the data through time $u-1$ that a participant is
at risk of the event being observed at time $u$; in other words,
$I_u=1$ means that all the variables
$R_0,L_1,R_1,L_2...,L_{u-1},R_{u-1}$ in the data vector (\ref{O})
equal $0$, which makes it possible that $L_u=1$.  Analogously, $J_u$
is the indicator based on the outcome data through time $u$ and
censoring data before time $u$ that a participant is at risk of
censoring at time $u$. By convention we let $J_0=1$.

The efficient influence function for estimation of $S(k, a)$
\cite[see][]{moore2009increasing} is equal to:
\begin{equation} 
D_\eta(O)= \sum_{u=1}^{k}I_u\times H_Y(u,A,W)\left\{L_u - m(u,a,W)\right\} +
  S(k,a,W) - S(k,a),\label{defD}
\end{equation}
where we have explicitly added the dependence of the auxiliary
covariate $H_Y$ on $(u,A,W)$ to the notation, and have denoted the
nuisance parameters with $\eta=(m, \pi_A, \pi_C)$. In what follows we will
use $\theta = S(k, a)$, and will use $\theta(\eta_1)$ to refer to the
target parameter evaluated at a specific distribution implied by
$\eta_1$. We will denote $\P f=\int f(o)\dd\P(o)$, and
$\P h(t,a,W) = \int h(t,a,w)\dd\P(w)$ for functions $f$ and $h$. The
efficient influence function has important implications for estimation
of $S(k,a)$. First, the variance of $D_\eta(O)$ is the non-parametric
efficiency bound, meaning that it is the smallest possible variance
achievable by any regular estimator \citep{Bickel97}. Second, the
efficient influence function characterizes the first order bias of a
plug-in estimator based on data-adaptive regression. Correction for
this first order bias will allow us to establish normality of the
estimators. Specifically, for any estimate $\widehat\eta$ we have the
following first order expansion around the true parameter value
$\theta(\eta)$, proved in Lemma~1 in the Supplementary materials of
\cite{diaz2018targeted}:
\begin{equation}
  \theta(\widehat\eta) - \theta(\eta) = -\P D_{\widehat\eta}+\rem_1(\widehat\eta),\label{eq:fo}
\end{equation}
where $\rem_1$ is a second order remainder term given by
{\small
\[\rem_1(\widehat\eta)= -\sum_{u=1}^k\int\frac{\widehat S(k, a,w)}{\widehat S(u, a,w)}S(u-1, a,w)\{m(u, a,w)-\widehat
  m(u, a,w)\}\left\{\frac{\pi_A(a,w) \Pi_C(u, a,w)}{\widehat \pi_A(a,w) \widehat
      \Pi_C(u, a,w)}- 1\right\}\dd\P(w),\]
}
  and $\theta(\widehat\eta)$ is the substitution estimator
  \[\theta(\widehat\eta)  = \frac{1}{n}\sum_{i=1}^n \prod_{u=1}^k\{1-\widehat
    m(u,a,W_i)\}.\] The following proposition establishing the
  robustness of $D_\eta$ to misspecification of the model $m$ will be
  useful to prove consistency of the estimator.
\begin{proposition}\label{propo:cons}
  Let $\eta_1=(m_1, \pi_{A,1}, \pi_{C,1})$ be such that either $m_1=m$ or
  $(\pi_{A,1}, \pi_{C,1})=(\pi_A, \pi_C)$. Then $\P D_{\eta_1} = 0$.
\end{proposition}
Recall the cross-fitting procedure described in the main document as
follows. Let ${\cal V}_1, \ldots, {\cal V}_J$ denote a
random partition of the index set $\{1, \ldots, n\}$ into $J$
prediction sets of approximately the same size. That is,
${\cal V}_j\subset \{1, \ldots, n\}$;
$\bigcup_{j=1}^J {\cal V}_j = \{1, \ldots, n\}$; and
${\cal V}_j\cap {\cal V}_{j'} = \emptyset$. In addition, for each $j$,
the associated training sample is given by
${\cal T}_j = \{1, \ldots, n\} \setminus {\cal V}_j$. Let $\widehat m_j$
denote the prediction algorithm trained in $\mathcal T_j$. Letting $j(i)$
denote the index of the validation set which contains observation $i$,
cross-fitting entails using only observations in $\mathcal T_{j(i)}$
for fitting models when making predictions about observation $i$. That
is, the outcome predictions for each subject $i$ are given by $\widehat
m_{j(i)}(u,a,W_i)$. Since only $\widehat m$ and not $(\widehat \pi_A,\widehat \pi_C)$ is
cross-fitted, we let $\widehat\eta_{j(i)}=(\widehat m_{j(i)}, \widehat \pi_A, \widehat
\pi_C)$ and $\widetilde\eta_{j(i)}=(\widetilde m_{j(i)}, \widehat \pi_A, \widehat
\pi_C)$. % We define the variance estimator as:
% \[\widetilde\sigma^2 = \frac{1}{n}\sum_{i=1}^n [D_{\widetilde\eta_{j(i)}}(O_i)]^2.\]

\section{Proof of Theorem \ref{theo:normal}}

In what follows we let $\Pnj$ denote the empirical distribution of the
prediction set ${\cal V}_j$, and let $\Gnj$ denote the associated
empirical process $\sqrt{n/J}(\Pnj-\P)$. Let $\Gn$ denote the
empirical process $\sqrt{n}(\Pn-\P)$.  We use $E(g(O_1,\ldots,O_n))$
to denote expectation with respect to the joint distribution of
$(O_1,\ldots,O_n)$, and use $a_n\lesssim b_n$ to mean $a_n\leq cb_n$
for universal constants $c$.  The following  lemmas will be useful
in the proof of the theorem.
\begin{lemma}\label{lemma:aux}
  Assume \ref{ass:randomcensor}, \ref{ass:ga}, and \ref{ass:gc} . Then
  we have $\Pi_C(k,a,w)$ does not depend on $w$, and $\pi_A(a,w)$ does not
  depend on $w$. Furthermore, we have
  \begin{align*}
    \sqrt{n}\{\widehat \Pi_C(k,a) - \Pi_C(k,a)\} &= \Gn \Delta_{k,a} +o_P(1),\\
    \sqrt{n}\{\widehat \pi_A(a) - \pi_A(a)\} &= \Gn \Lambda_a +o_P(1),
  \end{align*}
  for mean-zero functions $\Delta_{k,a}(O_i)$  and $\Lambda_a(O_i)$ of $(k,a)$ and $O_i$
  that do not depend on $W_i$.
\end{lemma}
\begin{proof}
  This lemma follows by application of the Delta method to the
  non-parametric maximum likelihood estimators $\widehat \pi_A$ and $\widehat \Pi_C$.
\end{proof}
\begin{lemma}\label{lemma:telescope}
  For two sequences $a_1,\ldots,a_m$ and $b_1,\ldots,b_m$ we have
  \[\prod_{t=1}^{m}(1-a_t) - \prod_{t=1}^{m}(1-b_t) = \sum_{t=1}^{m}\left\{\left[\prod_{k=1}^{t-1}(1-a_k)\right](b_t-a_t)\left[\prod_{k=t+1}^{m}(1-b_k)\right]\right\}.\]
\end{lemma}
\begin{proof}
  Replace $(b_t-a_t)$ by $(1 - a_t) - (1 - b_t)$ in the right hand
  side and expand the sum to notice it is a telescoping sum.
\end{proof}

The proof of Theorem~\ref{theo:normal} proceeds as follows.
\begin{proof}
  Since censoring is completely at random by \ref{ass:randomcensor},
  we have $\theta = S(k,a) = \int S(k,a,w)\dd\P(w)$.  Let
  $\widetilde\theta=\Stmle(k,a)$. Define $\sigma^2 = \var[D_{\eta_1}(O)]$, where
  $\eta_1=(m_1,\pi_A,\pi_C)$, and let
\begin{align*}
  \widetilde\Theta_n &= \sqrt{n}(\widetilde\theta - \theta)/\widetilde\sigma \\
  \check\Theta_n &= \sqrt{n}(\widetilde\theta - \theta)/\sigma \\
  \Theta_n &= \Gn D_{\eta_1} / \sigma.
\end{align*}
First, note that $\Theta_n\rightsquigarrow N(0,1)$ by the central
limit theorem. We will now show that $|\widetilde\Theta_n -
\Theta_n|=o_P(1)$, which would yield the result in the theorem. First,
note that
\begin{align*}
  |\widetilde\Theta_n - \Theta_n| &= |(\check\Theta_n -
  \Theta_n)(\sigma/\widetilde\sigma) +
                                \Theta_n(\sigma-\widetilde\sigma)/\widetilde\sigma|\\
                              &\leq |\check\Theta_n -
  \Theta_n|\,|\sigma/\widetilde\sigma| +
    |\Theta_n|\,|\sigma/\widetilde\sigma - 1|\\
                              &\lesssim |\check\Theta_n -
                                \Theta_n| + o_P(1),
\end{align*}
where the last inequality follows because $|\sigma/\widetilde\sigma -
1|=o_P(1)$ (which follows by Lemma~\ref{lemma:aux} and
\ref{ass:limitm}) and because $|\Theta_n|=O_P(1)$ by the central limit
theorem. We will now show that $|\check\Theta_n - \Theta_n|=o_P(1)$.

An application of (\ref{eq:fo}) with $\widehat\eta=\widetilde\eta$ yields
\begin{align*}
  \sqrt{n}(\widetilde\theta - \theta) &= -\sqrt{n}\P D_{\widetilde\eta} +
                                    \sqrt{n}\rem_1(\widetilde\eta)\\
                                  &= \sqrt{n}(\Pn-\P) D_{\widetilde\eta} +
                                    \sqrt{n}\rem_1(\widetilde\eta)\\
                                  &= \Gn D_{\eta_1} + \Gn(D_{\widetilde\eta}-D_{\eta_1}) +
                                    \sqrt{n}\rem_1(\widetilde\eta),
\end{align*}
where the second equality follows because $\Pn D_{\widetilde\eta} = 0$ by
definition of $\widetilde\eta$ \citep[see][]{diaz2019improved}. This implies
\[\check\Theta_n -
  \Theta_n = B_{n,2} + B_{n,1},\] where
$B_{n,2}=\Gn(D_{\widetilde\eta}-D_{\eta_1})$ and
$B_{n,1}=\sqrt{n}\rem_1(\widetilde\eta)$.

We first tackle the case of \ref{ass:m}.\ref{case:cross}, where the estimators for
$m$ are cross-fitted. Note that
\[  B_{n,2}  =\frac{1}{\sqrt{J}}\sum_{j=1}^J\Gnj(D_{\widetilde
    \eta_j} - D_{\eta_1}),
\]
and that $D_{\widetilde \eta_j}$ depends on the full sample through the
estimate of the parameter $\varepsilon$
of the logistic tilting model. To make this dependence explicit, we
introduce the notation
$D_{\widehat \eta_j,\widehat\varepsilon}=D_{\widetilde \eta_j}$. Let $\varepsilon_1$ denote the probability
limit of $\widehat\varepsilon$, which exists and is finite by Assumption A7. We can find a deterministic sequence
$\delta_n\to 0$ satisfying
$P(|\widehat\varepsilon - \varepsilon_1|<\delta_n)\to 1$. Let
${\cal F}_n^j=\{D_{\widehat \eta_j,\varepsilon} -
D_{\eta_1}:|\varepsilon - \varepsilon_1|<\delta_n\}$. Because the function $\widehat\eta_j$ is
fixed given the training data, we can apply Theorem 2.14.2 of
\cite{vanderVaart&Wellner96} to obtain
\begin{equation}
  E\left\{\sup_{f\in {\cal F}_n^j}|\Gnj f| \,\,\bigg|\,\, {\cal
      T}_j\right\}\lesssim \lVert F^j_n \rVert \int_0^1
  \sqrt{1+N_{[\,]}(\alpha \lVert F_n^j \rVert, {\cal F}_n^j,
    L_2(\P))}\dd\alpha,\label{eq:empproc}
\end{equation} where
$N_{[\,]}(\alpha \lVert F_n^j \rVert, {\cal F}_n^j, L_2(\P))$ is the
bracketing number and we take
$F_n^j=\sup_{\varepsilon: |\varepsilon-\varepsilon_1|<\delta_n} \lvert D_{\widehat \eta_j,\varepsilon} - D_{\eta_1} \rvert$
as an envelope for the class ${\cal F}_n^j$. Theorem 2.7.2 of \cite{vanderVaart&Wellner96} shows
\[\log N_{[\,]}(\alpha \lVert F_n^j \rVert, {\cal F}_n^j, L_2(\P))\lesssim
  \frac{1}{\alpha \lVert F_n^j \rVert}.\]
This shows
\begin{align*}
  \lVert F^j_n \rVert \int_0^1\sqrt{1+N_{[\,]}(\alpha
  \lVert F_n^j \rVert, {\cal F}_n^j, L_2(\P))}\dd\alpha &\lesssim
                                                          \int_0^1\sqrt{\lVert F^j_n \rVert^2+\frac{\lVert F^j_n
                                                          \rVert}{\alpha}}\dd\alpha\\ &\leq
                                                                                        \lVert F^j_n \rVert + \lVert F^j_n \rVert^{1/2}
                                                                                        \int_0^1\frac{1}{\alpha^{1/2}}\dd\alpha\\
                                                        &\leq \lVert F^j_n \rVert + 2 \lVert F^j_n \rVert^{1/2}.
\end{align*}
Since $D_{\widehat\eta_j,\widehat\varepsilon}\to D_{\eta_1}$ and $\delta_n\to 0$,
$\lVert F^j_n \rVert = o_P(1)$. The above argument  shows that 
$\sup_{f\in {\cal F}_n^j}|\Gnj f|=o_P(1)$ for each $j$, conditional on
${\cal T}_j$. Thus $|B_{n,2}|=o_P(1)$.

In the case of \ref{ass:m}.\ref{case:param}, where the estimators for
$m$ are \textit{not} cross-fitted but belong in a parametric family,
standard empirical process theory such as Example 19.7 of \cite{vanderVaart98}
shows that $D_{\widetilde\eta}$ takes values in a Donsker
class. Therefore, an application of Theorem 19.24 of \cite{vanderVaart98} yields $|B_{n,2}|=o_P(1)$.

We now show that $|B_{n,1}|=o_P(1)$. First, Lemma~\ref{lemma:aux}
along with the Delta method show that
\[\sqrt{n}\left\{\frac{\pi_A(a,w) \Pi_C(u, a,w)}{\widetilde \pi_A(a,w) \widetilde
      \Pi_C(u, a,w)}- 1\right\}=\Gn \Gamma_{k,a} + o_P(1),\]
for some function $\Gamma_{k,a}(O)$ not depending on $W$. Thus
\begin{align*}
  B_{n,1}&= -\sum_{u=1}^k\int\frac{\widetilde S(k,a,w)}{\widetilde S(u, a,w)}S(u-1, a,w)\{m(u, a,w)-\widetilde
           m(u, a,w)\}\left\{\Gn \Gamma_{k,a} + o_P(1)\right\}\dd\P(w)\\
         &= -\Gn \Gamma_{k,a} \sum_{u=1}^k\int\frac{\widetilde S(k,a,w)}{\widetilde S(u, a,w)}S(u-1, a,w)\{m(u, a,w)-\widetilde
           m(u, a,w)\}\dd\P(w) + o_P(1)\\
         &= \Gn \Gamma_{k,a} \int\{S(k, a,w)-\widetilde
           S(k, a,w)\}\dd\P(w) + o_P(1),
\end{align*}
where the last equality follows from
Lemma~\ref{lemma:telescope}. Expression (\ref{eq:fo}) together with
the assumptions of the theorem and Proposition~\ref{propo:cons} show that the estimator $\widetilde\theta$
is consistent, and thus
\[\int\{S(k, a,w)-\widetilde
S(k, a,w)\}\dd\P(w)=o_P(1).\]
The central limit theorem shows that $\Gn \Gamma_{k,a}=O_P(1)$, which
yields $|B_{n,1}| = o_P(1)$, concluding the proof of the theorem.
\end{proof}
\section{Tables with simulation results}

\subsection{Results for simulations with a positive effect and where the
  covariates are prognostic of the outcome}
\begin{table}[H]
  \centering
  \caption{Simulation results for the $\rmst$ of time to intubation or
    death at day 14 under a positive effect and covariates with prognostic
    power.}
  \label{tab:rmst:yy}
  \begin{tabular}{lccccccc}
    \toprule
    Estimator & $n$ & Effect size & P(Reject H$_0$) & $n\times\mse$ & $n\times \var$ & |Bias| & Rel. eff.\\ \midrule
    CF-MARS & 100 & 1.04 & 0.26 & 83.22 & 83.07 & 0.04 & 1.29 \\ 
    CF-RF & 100 & 1.04 & 0.24 & 71.13 & 71.13 & 0.00 & 1.10 \\ 
    CF-XGBoost & 100 & 1.04 & 0.23 & 70.33 & 70.33 & 0.00 & 1.09 \\ 
    $\ell_1$-LR & 100 & 1.04 & 0.28 & 58.49 & 58.47 & 0.02 & 0.90 \\ 
    LR & 100 & 1.04 & 0.34 & 66.61 & 66.61 & 0.00 & 1.03 \\ 
    MARS & 100 & 1.04 & 0.27 & 70.14 & 69.26 & 0.09 & 1.09 \\ 
    RF & 100 & 1.04 & 0.62 & 51.07 & 44.94 & 0.25 & 0.79 \\ 
    Unadjusted & 100 & 1.04 & 0.25 & 64.64 & 64.64 & 0.01 & 1.00 \\ 
    XGBoost & 100 & 1.04 & 0.27 & 64.07 & 64.06 & 0.01 & 0.99 \\ \addlinespace 
    CF-MARS & 500 & 1.04 & 0.85 & 61.36 & 61.35 & 0.00 & 0.93 \\ 
    CF-RF & 500 & 1.04 & 0.85 & 60.10 & 60.10 & 0.00 & 0.91 \\ 
    CF-XGBoost & 500 & 1.04 & 0.82 & 64.70 & 64.69 & 0.01 & 0.98 \\ 
    $\ell_1$-LR & 500 & 1.04 & 0.87 & 58.00 & 58.00 & 0.00 & 0.88 \\ 
    LR & 500 & 1.04 & 0.87 & 60.49 & 60.49 & 0.00 & 0.92 \\ 
    MARS & 500 & 1.04 & 0.86 & 58.01 & 57.95 & 0.01 & 0.88 \\ 
    RF & 500 & 1.04 & 0.97 & 66.02 & 55.76 & 0.14 & 1.00 \\ 
    Unadjusted & 500 & 1.04 & 0.82 & 65.85 & 65.85 & 0.00 & 1.00 \\ 
    XGBoost & 500 & 1.04 & 0.88 & 58.13 & 58.08 & 0.01 & 0.88 \\ \addlinespace 
    CF-MARS & 1500 & 1.04 & 1.00 & 60.51 & 60.50 & 0.00 & 0.93 \\ 
    CF-RF & 1500 & 1.04 & 1.00 & 47.88 & 47.88 & 0.00 & 0.74 \\ 
    CF-XGBoost & 1500 & 1.04 & 1.00 & 61.51 & 61.50 & 0.00 & 0.95 \\ 
    $\ell_1$-LR & 1500 & 1.04 & 1.00 & 58.15 & 58.14 & 0.00 & 0.89 \\ 
    LR & 1500 & 1.04 & 1.00 & 57.15 & 57.15 & 0.00 & 0.88 \\ 
    MARS & 1500 & 1.04 & 1.00 & 60.15 & 60.15 & 0.00 & 0.92 \\ 
    RF & 1500 & 1.04 & 1.00 & 61.29 & 50.61 & 0.08 & 0.94 \\ 
    Unadjusted & 1500 & 1.04 & 1.00 & 65.07 & 65.06 & 0.00 & 1.00 \\ 
    XGBoost & 1500 & 1.04 & 1.00 & 56.24 & 55.91 & 0.01 & 0.86 \\
    \bottomrule
  \end{tabular}
\end{table}

\begin{table}[H]
  \centering
  \caption{Simulation results for the $\rd$ of time to intubation or
    death at day 7 under a positive effect and covariates with prognostic
    power.}
  \label{tab:rd:yy}
  \begin{tabular}{lccccccc}
    \toprule
    Estimator & $n$ & Effect size & P(Reject H$_0$) & $n\times\mse$ & $n\times \var$ & |Bias| & Rel. eff.\\ \midrule
    CF-MARS & 100 & 0.10 & 0.27 & 0.74 & 0.73 & 0.00 & 1.23 \\ 
    CF-RF & 100 & 0.10 & 0.25 & 0.64 & 0.64 & 0.00 & 1.08 \\ 
    CF-XGBoost & 100 & 0.10 & 0.25 & 0.64 & 0.64 & 0.00 & 1.07 \\ 
    $\ell_1$-LR & 100 & 0.10 & 0.28 & 0.55 & 0.55 & 0.00 & 0.92 \\ 
    LR & 100 & 0.10 & 0.32 & 0.60 & 0.60 & 0.00 & 1.01 \\ 
    MARS & 100 & 0.10 & 0.28 & 0.64 & 0.63 & 0.01 & 1.07 \\ 
    RF & 100 & 0.10 & 0.50 & 0.56 & 0.52 & 0.02 & 0.94 \\ 
    Unadjusted & 100 & 0.10 & 0.26 & 0.60 & 0.60 & 0.00 & 1.00 \\ 
    XGBoost & 100 & 0.10 & 0.28 & 0.60 & 0.60 & 0.00 & 1.01 \\ \addlinespace 
    CF-MARS & 500 & 0.10 & 0.85 & 0.57 & 0.57 & 0.00 & 0.92 \\ 
    CF-RF & 500 & 0.10 & 0.85 & 0.56 & 0.56 & 0.00 & 0.91 \\ 
    CF-XGBoost & 500 & 0.10 & 0.84 & 0.59 & 0.59 & 0.00 & 0.96 \\ 
    $\ell_1$-LR & 500 & 0.10 & 0.87 & 0.54 & 0.54 & 0.00 & 0.89 \\ 
    LR & 500 & 0.10 & 0.86 & 0.57 & 0.57 & 0.00 & 0.92 \\ 
    MARS & 500 & 0.10 & 0.85 & 0.56 & 0.55 & 0.00 & 0.90 \\ 
    RF & 500 & 0.10 & 0.94 & 0.66 & 0.58 & 0.01 & 1.07 \\ 
    Unadjusted & 500 & 0.10 & 0.83 & 0.61 & 0.61 & 0.00 & 1.00 \\ 
    XGBoost & 500 & 0.10 & 0.87 & 0.55 & 0.55 & 0.00 & 0.90 \\ \addlinespace 
    CF-MARS & 1500 & 0.10 & 1.00 & 0.58 & 0.58 & 0.00 & 0.96 \\ 
    CF-RF & 1500 & 0.10 & 1.00 & 0.47 & 0.47 & 0.00 & 0.77 \\ 
    CF-XGBoost & 1500 & 0.10 & 1.00 & 0.56 & 0.56 & 0.00 & 0.94 \\ 
    $\ell_1$-LR & 1500 & 0.10 & 1.00 & 0.56 & 0.56 & 0.00 & 0.93 \\ 
    LR & 1500 & 0.10 & 1.00 & 0.55 & 0.55 & 0.00 & 0.92 \\ 
    MARS & 1500 & 0.10 & 1.00 & 0.57 & 0.57 & 0.00 & 0.94 \\ 
    RF & 1500 & 0.10 & 1.00 & 0.61 & 0.54 & 0.01 & 1.01 \\ 
    Unadjusted & 1500 & 0.10 & 1.00 & 0.60 & 0.60 & 0.00 & 1.00 \\ 
    XGBoost & 1500 & 0.10 & 1.00 & 0.54 & 0.53 & 0.00 & 0.89 \\
    \bottomrule
  \end{tabular}
\end{table}

\begin{table}[H]
  \centering
  \caption{Simulation results for the $\lor$ of the modified WHO scale
    at day 14 under a positive effect and covariates with prognostic
    power.}
  \label{tab:lor:yy}
  \begin{tabular}{lccccccc}
    \toprule
    Estimator & $n$ & Effect size & P(Reject H$_0$) & $n\times\mse$ & $n\times \var$ & |Bias| & Rel. eff.\\ \midrule
    CF-MARS & 100 & 0.60 & 0.35 &  73.00 &  70.25 & 0.17 & 0.13 \\ 
    CF-RF & 100 & 0.60 & 0.36 &  37.97 &  36.75 & 0.11 & 0.07 \\ 
    CF-XGBoost & 100 & 0.60 & 0.36 &  34.80 &  34.30 & 0.07 & 0.06 \\ 
    $\ell_1$-LR & 100 & 0.60 & 0.45 &  57.23 &  54.57 & 0.16 & 0.10 \\ 
    LR & 100 & 0.60 & 0.48 & 472.16 & 392.98 & 0.89 & 0.81 \\ 
    MARS & 100 & 0.60 & 0.45 & 192.05 & 176.11 & 0.40 & 0.33 \\ 
    RF & 100 & 0.60 & 0.48 &  32.02 &  32.01 & 0.01 & 0.05 \\ 
    Unadjusted & 100 & 0.60 & 0.42 & 582.27 & 441.29 & 1.19 & 1.00 \\ 
    XGBoost & 100 & 0.60 & 0.41 &  49.81 &  49.47 & 0.06 & 0.09 \\ \addlinespace 
    CF-MARS & 500 & 0.60 & 0.90 &  30.48 &  30.27 & 0.02 & 0.54 \\ 
    CF-RF & 500 & 0.60 & 0.93 &  16.15 &  16.13 & 0.01 & 0.29 \\ 
    CF-XGBoost & 500 & 0.60 & 0.92 &  18.88 &  18.86 & 0.01 & 0.34 \\ 
    $\ell_1$-LR & 500 & 0.60 & 0.93 &  18.74 &  18.68 & 0.01 & 0.33 \\ 
    LR & 500 & 0.60 & 0.92 &  32.86 &  32.53 & 0.03 & 0.59 \\ 
    MARS & 500 & 0.60 & 0.93 &  32.10 &  32.07 & 0.01 & 0.57 \\ 
    RF & 500 & 0.60 & 0.99 &  15.84 &  15.10 & 0.04 & 0.28 \\ 
    Unadjusted & 500 & 0.60 & 0.86 &  56.01 &  55.76 & 0.02 & 1.00 \\ 
    XGBoost & 500 & 0.60 & 0.95 &  16.29 &  15.42 & 0.04 & 0.29 \\ \addlinespace 
    CF-MARS & 1500 & 0.60 & 1.00 &  17.71 &  17.71 & 0.00 & 0.88 \\ 
    CF-RF & 1500 & 0.60 & 1.00 &  12.28 &  12.24 & 0.01 & 0.61 \\ 
    CF-XGBoost & 1500 & 0.60 & 1.00 &  15.28 &  15.27 & 0.00 & 0.76 \\ 
    $\ell_1$-LR & 1500 & 0.60 & 1.00 &  17.90 &  17.89 & 0.00 & 0.89 \\ 
    LR & 1500 & 0.60 & 1.00 &  17.12 &  17.04 & 0.01 & 0.85 \\ 
    MARS & 1500 & 0.60 & 1.00 &  16.72 &  16.71 & 0.00 & 0.83 \\ 
    RF & 1500 & 0.60 & 1.00 &   9.93 &   9.21 & 0.02 & 0.49 \\ 
    Unadjusted & 1500 & 0.60 & 1.00 &  20.20 &  20.20 & 0.00 & 1.00 \\ 
    XGBoost & 1500 & 0.60 & 1.00 &  11.21 &  10.81 & 0.02 & 0.55 \\
    \bottomrule
  \end{tabular}
\end{table}

\begin{table}[H]
  \centering
  \caption{Simulation results for the $\mw$ estimand of the modified
    WHO scale at day 14 under a positive effect and covariates with prognostic
    power.}
  \label{tab:mw:yy}
  \begin{tabular}{lccccccc}
    \toprule
    Estimator & $n$ & Effect size & P(Reject H$_0$) & $n\times\mse$ & $n\times \var$ & |Bias| & Rel. eff.\\ \midrule
    CF-MARS & 100 & 0.46 & 0.24 & 0.66 & 0.66 & 0.00 & 2.52 \\ 
    CF-RF & 100 & 0.46 & 0.17 & 0.24 & 0.24 & 0.00 & 0.90 \\ 
    CF-XGBoost & 100 & 0.46 & 0.15 & 0.24 & 0.24 & 0.00 & 0.92 \\ 
    $\ell_1$-LR & 100 & 0.46 & 0.18 & 0.23 & 0.23 & 0.00 & 0.86 \\ 
    LR & 100 & 0.46 & 0.45 & 1.68 & 1.66 & 0.01 & 6.40 \\ 
    MARS & 100 & 0.46 & 0.23 & 0.49 & 0.49 & 0.00 & 1.85 \\ 
    RF & 100 & 0.46 & 0.22 & 0.20 & 0.19 & 0.01 & 0.76 \\ 
    Unadjusted & 100 & 0.46 & 0.14 & 0.26 & 0.26 & 0.00 & 1.00 \\ 
    XGBoost & 100 & 0.46 & 0.19 & 0.21 & 0.21 & 0.00 & 0.81 \\ \addlinespace 
    CF-MARS & 500 & 0.46 & 0.51 & 0.52 & 0.51 & 0.01 & 2.01 \\ 
    CF-RF & 500 & 0.46 & 0.56 & 0.20 & 0.20 & 0.00 & 0.79 \\ 
    CF-XGBoost & 500 & 0.46 & 0.53 & 0.23 & 0.22 & 0.00 & 0.87 \\ 
    $\ell_1$-LR & 500 & 0.46 & 0.53 & 0.21 & 0.21 & 0.00 & 0.82 \\ 
    LR & 500 & 0.46 & 0.54 & 0.24 & 0.24 & 0.00 & 0.92 \\ 
    MARS & 500 & 0.46 & 0.51 & 0.26 & 0.26 & 0.00 & 1.02 \\ 
    RF & 500 & 0.46 & 0.76 & 0.14 & 0.14 & 0.00 & 0.55 \\ 
    Unadjusted & 500 & 0.46 & 0.44 & 0.26 & 0.26 & 0.00 & 1.00 \\ 
    XGBoost & 500 & 0.46 & 0.59 & 0.20 & 0.20 & 0.00 & 0.76 \\ \addlinespace 
    CF-MARS & 1500 & 0.46 & 0.93 & 0.23 & 0.22 & 0.00 & 0.86 \\ 
    CF-RF & 1500 & 0.46 & 0.99 & 0.15 & 0.14 & 0.00 & 0.57 \\ 
    CF-XGBoost & 1500 & 0.46 & 0.97 & 0.19 & 0.18 & 0.00 & 0.73 \\ 
    $\ell_1$-LR & 1500 & 0.46 & 0.93 & 0.23 & 0.22 & 0.00 & 0.86 \\ 
    LR & 1500 & 0.46 & 0.94 & 0.23 & 0.21 & 0.00 & 0.85 \\ 
    MARS & 1500 & 0.46 & 0.94 & 0.21 & 0.21 & 0.00 & 0.80 \\ 
    RF & 1500 & 0.46 & 1.00 & 0.11 & 0.11 & 0.00 & 0.40 \\ 
    Unadjusted & 1500 & 0.46 & 0.88 & 0.27 & 0.26 & 0.00 & 1.00 \\ 
    XGBoost & 1500 & 0.46 & 0.99 & 0.14 & 0.13 & 0.00 & 0.52 \\
    \bottomrule
  \end{tabular}
\end{table}

\subsection{Results for simulations with null treatment effect and where the
  covariates are prognostic of the outcome}
\begin{table}[H]
  \centering
  \caption{Simulation results for the $\rmst$ of time to intubation or
    death at day 14 under null treatment effect and covariates with prognostic
    power.}
  \label{tab:rmst:ny}
  \begin{tabular}{lccccccc}
    \toprule
    Estimator & $n$ & Effect size & P(Reject H$_0$) & $n\times\mse$ & $n\times \var$ & |Bias| & Rel. eff.\\ \midrule
    CF-MARS & 100 & 0.00 & 0.10 & 126.72 & 121.56 & 0.23 & 1.44 \\ 
    CF-RF & 100 & 0.00 & 0.05 &  88.87 &  88.81 & 0.02 & 1.01 \\ 
    CF-XGBoost & 100 & 0.00 & 0.05 &  90.87 &  90.84 & 0.02 & 1.03 \\ 
    $\ell_1$-LR & 100 & 0.00 & 0.07 &  82.16 &  82.14 & 0.01 & 0.93 \\ 
    LR & 100 & 0.00 & 0.09 &  86.83 &  86.82 & 0.01 & 0.99 \\ 
    MARS & 100 & 0.00 & 0.09 &  88.35 &  87.88 & 0.07 & 1.00 \\ 
    RF & 100 & 0.00 & 0.30 &  51.79 &  51.78 & 0.01 & 0.59 \\ 
    Unadjusted & 100 & 0.00 & 0.06 &  87.92 &  87.85 & 0.03 & 1.00 \\ 
    XGBoost & 100 & 0.00 & 0.05 &  78.09 &  78.09 & 0.01 & 0.89 \\ \addlinespace 
    CF-MARS & 500 & 0.00 & 0.05 &  80.81 &  80.76 & 0.01 & 0.95 \\ 
    CF-RF & 500 & 0.00 & 0.05 &  75.68 &  75.68 & 0.00 & 0.89 \\ 
    CF-XGBoost & 500 & 0.00 & 0.05 &  83.71 &  83.70 & 0.00 & 0.98 \\ 
    $\ell_1$-LR & 500 & 0.00 & 0.05 &  73.05 &  73.03 & 0.01 & 0.86 \\ 
    LR & 500 & 0.00 & 0.06 &  79.64 &  79.63 & 0.00 & 0.94 \\ 
    MARS & 500 & 0.00 & 0.05 &  78.44 &  78.44 & 0.00 & 0.92 \\ 
    RF & 500 & 0.00 & 0.29 &  48.37 &  48.37 & 0.00 & 0.57 \\ 
    Unadjusted & 500 & 0.00 & 0.05 &  85.12 &  85.10 & 0.01 & 1.00 \\ 
    XGBoost & 500 & 0.00 & 0.07 &  72.98 &  72.98 & 0.00 & 0.86 \\ \addlinespace 
    CF-MARS & 1500 & 0.00 & 0.05 &  73.18 &  73.18 & 0.00 & 0.85 \\ 
    CF-RF & 1500 & 0.00 & 0.05 &  56.36 &  56.35 & 0.00 & 0.66 \\ 
    CF-XGBoost & 1500 & 0.00 & 0.05 &  80.36 &  80.36 & 0.00 & 0.94 \\ 
    $\ell_1$-LR & 1500 & 0.00 & 0.05 &  75.52 &  75.51 & 0.00 & 0.88 \\ 
    LR & 1500 & 0.00 & 0.06 &  75.39 &  75.38 & 0.00 & 0.88 \\ 
    MARS & 1500 & 0.00 & 0.05 &  76.55 &  76.51 & 0.00 & 0.89 \\ 
    RF & 1500 & 0.00 & 0.30 &  27.65 &  27.64 & 0.00 & 0.32 \\ 
    Unadjusted & 1500 & 0.00 & 0.05 &  85.87 &  85.83 & 0.00 & 1.00 \\ 
    XGBoost & 1500 & 0.00 & 0.07 &  64.40 &  64.40 & 0.00 & 0.75 \\
    \bottomrule
  \end{tabular}
\end{table}

\begin{table}[H]
  \centering
  \caption{Simulation results for the $\rd$ of time to intubation or
    death at day 7 under null treatment effect and covariates with prognostic
    power.}
  \label{tab:rd:ny}
  \begin{tabular}{lccccccc}
    \toprule
    Estimator & $n$ & Effect size & P(Reject H$_0$) & $n\times\mse$ & $n\times \var$ & |Bias| & Rel. eff.\\ \midrule
    CF-MARS & 100 & 0.00 & 0.09 & 1.00 & 0.96 & 0.02 & 1.36 \\ 
    CF-RF & 100 & 0.00 & 0.05 & 0.75 & 0.75 & 0.00 & 1.02 \\ 
    CF-XGBoost & 100 & 0.00 & 0.05 & 0.75 & 0.75 & 0.00 & 1.03 \\ 
    $\ell_1$-LR & 100 & 0.00 & 0.07 & 0.69 & 0.69 & 0.00 & 0.95 \\ 
    LR & 100 & 0.00 & 0.09 & 0.72 & 0.72 & 0.00 & 0.98 \\ 
    MARS & 100 & 0.00 & 0.08 & 0.72 & 0.72 & 0.01 & 0.99 \\ 
    RF & 100 & 0.00 & 0.26 & 0.51 & 0.51 & 0.00 & 0.70 \\ 
    Unadjusted & 100 & 0.00 & 0.06 & 0.73 & 0.73 & 0.00 & 1.00 \\ 
    XGBoost & 100 & 0.00 & 0.05 & 0.66 & 0.66 & 0.00 & 0.90 \\ \addlinespace 
    CF-MARS & 500 & 0.00 & 0.05 & 0.69 & 0.68 & 0.00 & 0.95 \\ 
    CF-RF & 500 & 0.00 & 0.04 & 0.63 & 0.63 & 0.00 & 0.88 \\ 
    CF-XGBoost & 500 & 0.00 & 0.04 & 0.70 & 0.70 & 0.00 & 0.98 \\ 
    $\ell_1$-LR & 500 & 0.00 & 0.05 & 0.63 & 0.63 & 0.00 & 0.88 \\ 
    LR & 500 & 0.00 & 0.06 & 0.69 & 0.69 & 0.00 & 0.95 \\ 
    MARS & 500 & 0.00 & 0.06 & 0.68 & 0.68 & 0.00 & 0.94 \\ 
    RF & 500 & 0.00 & 0.25 & 0.45 & 0.45 & 0.00 & 0.62 \\ 
    Unadjusted & 500 & 0.00 & 0.05 & 0.72 & 0.72 & 0.00 & 1.00 \\ 
    XGBoost & 500 & 0.00 & 0.06 & 0.62 & 0.62 & 0.00 & 0.86 \\ \addlinespace 
    CF-MARS & 1500 & 0.00 & 0.04 & 0.62 & 0.62 & 0.00 & 0.86 \\ 
    CF-RF & 1500 & 0.00 & 0.05 & 0.49 & 0.49 & 0.00 & 0.67 \\ 
    CF-XGBoost & 1500 & 0.00 & 0.05 & 0.68 & 0.68 & 0.00 & 0.94 \\ 
    $\ell_1$-LR & 1500 & 0.00 & 0.05 & 0.64 & 0.64 & 0.00 & 0.89 \\ 
    LR & 1500 & 0.00 & 0.05 & 0.65 & 0.65 & 0.00 & 0.90 \\ 
    MARS & 1500 & 0.00 & 0.05 & 0.65 & 0.65 & 0.00 & 0.90 \\ 
    RF & 1500 & 0.00 & 0.25 & 0.26 & 0.26 & 0.00 & 0.36 \\ 
    Unadjusted & 1500 & 0.00 & 0.05 & 0.72 & 0.72 & 0.00 & 1.00 \\ 
    XGBoost & 1500 & 0.00 & 0.07 & 0.56 & 0.56 & 0.00 & 0.78 \\
    \bottomrule
  \end{tabular}
\end{table}

\begin{table}[H]
  \centering
  \caption{Simulation results for the $\lor$ of the modified WHO scale
    at day 14 under null treatment effect and covariates with prognostic
    power.}
  \label{tab:lor:ny}
  \begin{tabular}{lccccccc}
    \toprule
    Estimator & $n$ & Effect size & P(Reject H$_0$) & $n\times\mse$ & $n\times \var$ & |Bias| & Rel. eff.\\ \midrule
    CF-MARS & 100 & 0.00 & 0.07 &  21.27 &  21.27 & 0.00 & 0.55 \\ 
    CF-RF & 100 & 0.00 & 0.06 &  16.93 &  16.93 & 0.01 & 0.44 \\ 
    CF-XGBoost & 100 & 0.00 & 0.07 &  18.12 &  18.12 & 0.00 & 0.47 \\ 
    $\ell_1$-LR & 100 & 0.00 & 0.09 &  28.01 &  28.01 & 0.00 & 0.73 \\ 
    LR & 100 & 0.00 & 0.11 & 111.68 & 111.61 & 0.03 & 2.89 \\ 
    MARS & 100 & 0.00 & 0.09 &  60.09 &  60.08 & 0.01 & 1.56 \\ 
    RF & 100 & 0.00 & 0.13 &  21.91 &  21.90 & 0.01 & 0.57 \\ 
    Unadjusted & 100 & 0.00 & 0.06 &  38.60 &  38.60 & 0.00 & 1.00 \\ 
    XGBoost & 100 & 0.00 & 0.10 &  18.97 &  18.96 & 0.01 & 0.49 \\ \addlinespace 
    CF-MARS & 500 & 0.00 & 0.05 &  14.70 &  14.70 & 0.00 & 0.88 \\ 
    CF-RF & 500 & 0.00 & 0.05 &  12.57 &  12.57 & 0.00 & 0.75 \\ 
    CF-XGBoost & 500 & 0.00 & 0.06 &  14.07 &  14.06 & 0.00 & 0.84 \\ 
    $\ell_1$-LR & 500 & 0.00 & 0.06 &  13.51 &  13.51 & 0.00 & 0.81 \\ 
    LR & 500 & 0.00 & 0.06 &  13.76 &  13.75 & 0.00 & 0.82 \\ 
    MARS & 500 & 0.00 & 0.05 &  12.75 &  12.75 & 0.00 & 0.76 \\ 
    RF & 500 & 0.00 & 0.14 &   7.34 &   7.34 & 0.00 & 0.44 \\ 
    Unadjusted & 500 & 0.00 & 0.05 &  16.74 &  16.74 & 0.00 & 1.00 \\ 
    XGBoost & 500 & 0.00 & 0.08 &  11.07 &  11.07 & 0.00 & 0.66 \\ \addlinespace 
    CF-MARS & 1500 & 0.00 & 0.05 &  13.35 &  13.35 & 0.00 & 0.82 \\ 
    CF-RF & 1500 & 0.00 & 0.05 &   8.36 &   8.36 & 0.00 & 0.51 \\ 
    CF-XGBoost & 1500 & 0.00 & 0.05 &  11.03 &  11.03 & 0.00 & 0.68 \\ 
    $\ell_1$-LR & 1500 & 0.00 & 0.05 &  12.87 &  12.87 & 0.00 & 0.79 \\ 
    LR & 1500 & 0.00 & 0.05 &  13.27 &  13.27 & 0.00 & 0.81 \\ 
    MARS & 1500 & 0.00 & 0.05 &  12.52 &  12.52 & 0.00 & 0.77 \\ 
    RF & 1500 & 0.00 & 0.15 &   4.56 &   4.56 & 0.00 & 0.28 \\ 
    Unadjusted & 1500 & 0.00 & 0.05 &  16.30 &  16.29 & 0.00 & 1.00 \\ 
    XGBoost & 1500 & 0.00 & 0.16 &   6.37 &   6.37 & 0.00 & 0.39 \\
    \bottomrule
  \end{tabular}
\end{table}

\begin{table}[H]
  \centering
  \caption{Simulation results for the $\mw$ estimand of the modified WHO scale
    at day 14 under null treatment effect and covariates with prognostic
    power.}
  \label{tab:mw:ny}
  \begin{tabular}{lccccccc}
    \toprule
    Estimator & $n$ & Effect size & P(Reject H$_0$) & $n\times\mse$ & $n\times \var$ & |Bias| & Rel. eff.\\ \midrule
    CF-MARS & 100 & 0.50 & 0.11 & 0.54 & 0.54 & 0.00 & 1.95 \\ 
    CF-RF & 100 & 0.50 & 0.07 & 0.23 & 0.23 & 0.00 & 0.84 \\ 
    CF-XGBoost & 100 & 0.50 & 0.07 & 0.25 & 0.25 & 0.00 & 0.92 \\ 
    $\ell_1$-LR & 100 & 0.50 & 0.07 & 0.22 & 0.22 & 0.00 & 0.80 \\ 
    LR & 100 & 0.50 & 0.48 & 2.14 & 2.14 & 0.00 & 7.77 \\ 
    MARS & 100 & 0.50 & 0.12 & 0.43 & 0.43 & 0.00 & 1.58 \\ 
    RF & 100 & 0.50 & 0.13 & 0.16 & 0.16 & 0.00 & 0.57 \\ 
    Unadjusted & 100 & 0.50 & 0.06 & 0.28 & 0.28 & 0.00 & 1.00 \\ 
    XGBoost & 100 & 0.50 & 0.09 & 0.21 & 0.21 & 0.00 & 0.75 \\ \addlinespace 
    CF-MARS & 500 & 0.50 & 0.06 & 0.28 & 0.28 & 0.00 & 1.07 \\ 
    CF-RF & 500 & 0.50 & 0.05 & 0.20 & 0.20 & 0.00 & 0.74 \\ 
    CF-XGBoost & 500 & 0.50 & 0.06 & 0.22 & 0.22 & 0.00 & 0.85 \\ 
    $\ell_1$-LR & 500 & 0.50 & 0.05 & 0.21 & 0.21 & 0.00 & 0.80 \\ 
    LR & 500 & 0.50 & 0.08 & 0.79 & 0.79 & 0.00 & 2.98 \\ 
    MARS & 500 & 0.50 & 0.06 & 0.27 & 0.27 & 0.00 & 1.01 \\ 
    RF & 500 & 0.50 & 0.14 & 0.13 & 0.13 & 0.00 & 0.48 \\ 
    Unadjusted & 500 & 0.50 & 0.05 & 0.27 & 0.27 & 0.00 & 1.00 \\ 
    XGBoost & 500 & 0.50 & 0.08 & 0.18 & 0.18 & 0.00 & 0.69 \\ \addlinespace 
    CF-MARS & 1500 & 0.50 & 0.05 & 0.22 & 0.22 & 0.00 & 0.84 \\ 
    CF-RF & 1500 & 0.50 & 0.05 & 0.13 & 0.13 & 0.00 & 0.52 \\ 
    CF-XGBoost & 1500 & 0.50 & 0.06 & 0.18 & 0.18 & 0.00 & 0.71 \\ 
    $\ell_1$-LR & 1500 & 0.50 & 0.05 & 0.21 & 0.21 & 0.00 & 0.81 \\ 
    LR & 1500 & 0.50 & 0.06 & 0.22 & 0.22 & 0.00 & 0.84 \\ 
    MARS & 1500 & 0.50 & 0.05 & 0.22 & 0.22 & 0.00 & 0.84 \\ 
    RF & 1500 & 0.50 & 0.16 & 0.20 & 0.20 & 0.00 & 0.78 \\ 
    Unadjusted & 1500 & 0.50 & 0.05 & 0.26 & 0.26 & 0.00 & 1.00 \\ 
    XGBoost & 1500 & 0.50 & 0.15 & 0.11 & 0.11 & 0.00 & 0.43 \\
    \bottomrule
  \end{tabular}
\end{table}

\subsection{Results for simulations with a positive effect and where the
  covariates are not prognostic of the outcome}

\begin{table}[H]
  \centering
  \caption{Simulation results for the $\rmst$ of the time to
    intubation or death at day 14 under a positive effect and covariates
    with no prognostic power.}
  \label{tab:rmst:yn}
  \begin{tabular}{lccccccc}
    \toprule
    Estimator & $n$ & Effect size & P(Reject H$_0$) & $n\times\mse$ & $n\times \var$ & |Bias| & Rel. eff.\\ \midrule
    CF-MARS & 100 & 1.04 & 0.25 & 94.58 & 94.56 & 0.01 & 1.46 \\ 
    CF-RF & 100 & 1.04 & 0.21 & 77.13 & 77.13 & 0.00 & 1.19 \\ 
    CF-XGBoost & 100 & 1.04 & 0.23 & 77.43 & 77.41 & 0.01 & 1.20 \\ 
    $\ell_1$-LR & 100 & 1.04 & 0.24 & 64.21 & 64.21 & 0.01 & 0.99 \\ 
    LR & 100 & 1.04 & 0.30 & 75.60 & 75.55 & 0.02 & 1.17 \\ 
    MARS & 100 & 1.04 & 0.25 & 73.38 & 72.61 & 0.09 & 1.14 \\ 
    RF & 100 & 1.04 & 0.58 & 68.20 & 64.12 & 0.20 & 1.06 \\ 
    Unadjusted & 100 & 1.04 & 0.25 & 64.64 & 64.64 & 0.01 & 1.00 \\ 
    XGBoost & 100 & 1.04 & 0.26 & 66.74 & 66.74 & 0.01 & 1.03 \\ \addlinespace 
    CF-MARS & 500 & 1.04 & 0.81 & 67.36 & 67.36 & 0.00 & 1.02 \\ 
    CF-RF & 500 & 1.04 & 0.77 & 74.08 & 74.06 & 0.01 & 1.12 \\ 
    CF-XGBoost & 500 & 1.04 & 0.77 & 73.22 & 73.21 & 0.00 & 1.11 \\ 
    $\ell_1$-LR & 500 & 1.04 & 0.83 & 66.61 & 66.61 & 0.00 & 1.01 \\ 
    LR & 500 & 1.04 & 0.83 & 68.06 & 68.04 & 0.01 & 1.03 \\ 
    MARS & 500 & 1.04 & 0.82 & 64.56 & 64.54 & 0.01 & 0.98 \\ 
    RF & 500 & 1.04 & 0.91 & 81.56 & 79.70 & 0.06 & 1.24 \\ 
    Unadjusted & 500 & 1.04 & 0.82 & 65.85 & 65.85 & 0.00 & 1.00 \\ 
    XGBoost & 500 & 1.04 & 0.83 & 63.75 & 63.70 & 0.01 & 0.97 \\ \addlinespace 
    CF-MARS & 1500 & 1.04 & 1.00 & 65.62 & 65.61 & 0.00 & 1.01 \\ 
    CF-RF & 1500 & 1.04 & 1.00 & 73.92 & 73.91 & 0.00 & 1.14 \\ 
    CF-XGBoost & 1500 & 1.04 & 1.00 & 70.56 & 70.55 & 0.00 & 1.08 \\ 
    $\ell_1$-LR & 1500 & 1.04 & 1.00 & 65.55 & 65.54 & 0.00 & 1.01 \\ 
    LR & 1500 & 1.04 & 1.00 & 65.55 & 65.55 & 0.00 & 1.01 \\ 
    MARS & 1500 & 1.04 & 1.00 & 65.68 & 65.67 & 0.00 & 1.01 \\ 
    RF & 1500 & 1.04 & 1.00 & 74.73 & 74.72 & 0.00 & 1.15 \\ 
    Unadjusted & 1500 & 1.04 & 1.00 & 65.07 & 65.06 & 0.00 & 1.00 \\ 
    XGBoost & 1500 & 1.04 & 1.00 & 66.71 & 66.69 & 0.00 & 1.03 \\
    \bottomrule
  \end{tabular}
\end{table}

\begin{table}[H]
  \centering
  \caption{Simulation results for the $\rd$ of the time to
    intubation or death at day 7 under a positive effect and covariates
    with no prognostic power.}
  \label{tab:rd:yn}
  \begin{tabular}{lccccccc}
    \toprule
    Estimator & $n$ & Effect size & P(Reject H$_0$) & $n\times\mse$ & $n\times \var$ & |Bias| & Rel. eff.\\ \midrule
    CF-MARS & 100 & 0.10 & 0.26 & 0.83 & 0.83 & 0.00 & 1.39 \\ 
    CF-RF & 100 & 0.10 & 0.22 & 0.68 & 0.68 & 0.00 & 1.14 \\ 
    CF-XGBoost & 100 & 0.10 & 0.26 & 0.70 & 0.70 & 0.00 & 1.17 \\ 
    $\ell_1$-LR & 100 & 0.10 & 0.26 & 0.60 & 0.60 & 0.00 & 1.00 \\ 
    LR & 100 & 0.10 & 0.30 & 0.67 & 0.67 & 0.00 & 1.13 \\ 
    MARS & 100 & 0.10 & 0.27 & 0.67 & 0.67 & 0.01 & 1.13 \\ 
    RF & 100 & 0.10 & 0.51 & 0.71 & 0.68 & 0.02 & 1.18 \\ 
    Unadjusted & 100 & 0.10 & 0.26 & 0.60 & 0.60 & 0.00 & 1.00 \\ 
    XGBoost & 100 & 0.10 & 0.27 & 0.62 & 0.62 & 0.00 & 1.03 \\ \addlinespace 
    CF-MARS & 500 & 0.10 & 0.82 & 0.61 & 0.61 & 0.00 & 1.00 \\ 
    CF-RF & 500 & 0.10 & 0.79 & 0.66 & 0.66 & 0.00 & 1.08 \\ 
    CF-XGBoost & 500 & 0.10 & 0.79 & 0.66 & 0.66 & 0.00 & 1.07 \\ 
    $\ell_1$-LR & 500 & 0.10 & 0.83 & 0.61 & 0.61 & 0.00 & 0.99 \\ 
    LR & 500 & 0.10 & 0.83 & 0.62 & 0.62 & 0.00 & 1.02 \\ 
    MARS & 500 & 0.10 & 0.83 & 0.59 & 0.59 & 0.00 & 0.96 \\ 
    RF & 500 & 0.10 & 0.89 & 0.79 & 0.77 & 0.01 & 1.28 \\ 
    Unadjusted & 500 & 0.10 & 0.83 & 0.61 & 0.61 & 0.00 & 1.00 \\ 
    XGBoost & 500 & 0.10 & 0.83 & 0.60 & 0.59 & 0.00 & 0.97 \\ \addlinespace 
    CF-MARS & 1500 & 0.10 & 1.00 & 0.61 & 0.61 & 0.00 & 1.01 \\ 
    CF-RF & 1500 & 0.10 & 1.00 & 0.67 & 0.67 & 0.00 & 1.11 \\ 
    CF-XGBoost & 1500 & 0.10 & 1.00 & 0.63 & 0.63 & 0.00 & 1.05 \\ 
    $\ell_1$-LR & 1500 & 0.10 & 1.00 & 0.60 & 0.60 & 0.00 & 0.99 \\ 
    LR & 1500 & 0.10 & 1.00 & 0.60 & 0.60 & 0.00 & 0.99 \\ 
    MARS & 1500 & 0.10 & 1.00 & 0.60 & 0.60 & 0.00 & 0.99 \\ 
    RF & 1500 & 0.10 & 1.00 & 0.69 & 0.69 & 0.00 & 1.14 \\ 
    Unadjusted & 1500 & 0.10 & 1.00 & 0.60 & 0.60 & 0.00 & 1.00 \\ 
    XGBoost & 1500 & 0.10 & 1.00 & 0.61 & 0.61 & 0.00 & 1.01 \\
    \bottomrule
  \end{tabular}
\end{table}

\begin{table}[H]
  \centering
  \caption{Simulation results for the $\lor$ of the modified WHO scale
    at day 14 under a positive effect and covariates with no
    prognostic power.}
  \label{tab:lor:yn}
  \begin{tabular}{lccccccc}
    \toprule
    Estimator & $n$ & Effect size & P(Reject H$_0$) & $n\times\mse$ & $n\times \var$ & |Bias| & Rel. eff.\\ \midrule
    CF-MARS & 100 & 0.60 & 0.35 &  67.15 &  64.59 & 0.16 & 0.12 \\ 
    CF-RF & 100 & 0.60 & 0.37 &  62.46 &  58.98 & 0.19 & 0.11 \\ 
    CF-XGBoost & 100 & 0.60 & 0.40 &  76.06 &  72.37 & 0.19 & 0.13 \\ 
    $\ell_1$-LR & 100 & 0.60 & 0.44 &  95.99 &  90.49 & 0.23 & 0.16 \\ 
    LR & 100 & 0.60 & 0.41 & 392.58 & 330.71 & 0.79 & 0.67 \\ 
    MARS & 100 & 0.60 & 0.41 & 187.85 & 167.51 & 0.45 & 0.32 \\ 
    RF & 100 & 0.60 & 0.45 &  61.04 &  59.51 & 0.12 & 0.10 \\ 
    Unadjusted & 100 & 0.60 & 0.42 & 582.27 & 441.29 & 1.19 & 1.00 \\ 
    XGBoost & 100 & 0.60 & 0.43 & 132.91 & 124.69 & 0.29 & 0.23 \\ \addlinespace 
    CF-MARS & 500 & 0.60 & 0.87 &  31.28 &  31.16 & 0.02 & 0.56 \\ 
    CF-RF & 500 & 0.60 & 0.87 &  23.06 &  23.05 & 0.00 & 0.41 \\ 
    CF-XGBoost & 500 & 0.60 & 0.88 &  25.83 &  25.62 & 0.02 & 0.46 \\ 
    $\ell_1$-LR & 500 & 0.60 & 0.87 &  24.01 &  23.98 & 0.01 & 0.43 \\ 
    LR & 500 & 0.60 & 0.86 &  53.78 &  53.45 & 0.03 & 0.96 \\ 
    MARS & 500 & 0.60 & 0.87 &  39.20 &  39.04 & 0.02 & 0.70 \\ 
    RF & 500 & 0.60 & 0.91 &  33.77 &  33.77 & 0.00 & 0.60 \\ 
    Unadjusted & 500 & 0.60 & 0.86 &  56.01 &  55.76 & 0.02 & 1.00 \\ 
    XGBoost & 500 & 0.60 & 0.89 &  22.54 &  22.54 & 0.00 & 0.40 \\ \addlinespace 
    CF-MARS & 1500 & 0.60 & 1.00 &  20.98 &  20.98 & 0.00 & 1.04 \\ 
    CF-RF & 1500 & 0.60 & 1.00 &  20.93 &  20.92 & 0.00 & 1.04 \\ 
    CF-XGBoost & 1500 & 0.60 & 1.00 &  22.14 &  22.09 & 0.01 & 1.10 \\ 
    $\ell_1$-LR & 1500 & 0.60 & 1.00 &  20.81 &  20.80 & 0.00 & 1.03 \\ 
    LR & 1500 & 0.60 & 1.00 &  20.27 &  20.26 & 0.00 & 1.00 \\ 
    MARS & 1500 & 0.60 & 1.00 &  20.16 &  20.15 & 0.00 & 1.00 \\ 
    RF & 1500 & 0.60 & 1.00 &  20.51 &  20.48 & 0.00 & 1.02 \\ 
    Unadjusted & 1500 & 0.60 & 1.00 &  20.20 &  20.20 & 0.00 & 1.00 \\ 
    XGBoost & 1500 & 0.60 & 1.00 &  20.48 &  20.44 & 0.01 & 1.01 \\
    \bottomrule
  \end{tabular}
\end{table}

\begin{table}[H]
  \centering
  \caption{Simulation results for the $\mw$ of the modified WHO scale
    at day 14 under a positive effect and covariates with no
    prognostic power.}
  \label{tab:mw:yn}
  \begin{tabular}{lccccccc}
    \toprule
    Estimator & $n$ & Effect size & P(Reject H$_0$) & $n\times\mse$ & $n\times \var$ & |Bias| & Rel. eff.\\ \midrule
    CF-MARS & 100 & 0.46 & 0.18 & 0.54 & 0.54 & 0.00 & 2.05 \\ 
    CF-RF & 100 & 0.46 & 0.15 & 0.29 & 0.29 & 0.00 & 1.11 \\ 
    CF-XGBoost & 100 & 0.46 & 0.15 & 0.29 & 0.29 & 0.00 & 1.10 \\ 
    $\ell_1$-LR & 100 & 0.46 & 0.16 & 0.26 & 0.26 & 0.00 & 1.00 \\ 
    LR & 100 & 0.46 & 0.44 & 1.69 & 1.67 & 0.01 & 6.43 \\ 
    MARS & 100 & 0.46 & 0.18 & 0.43 & 0.43 & 0.00 & 1.63 \\ 
    RF & 100 & 0.46 & 0.19 & 0.24 & 0.23 & 0.01 & 0.91 \\ 
    Unadjusted & 100 & 0.46 & 0.14 & 0.26 & 0.26 & 0.00 & 1.00 \\ 
    XGBoost & 100 & 0.46 & 0.18 & 0.26 & 0.26 & 0.00 & 1.00 \\ \addlinespace 
    CF-MARS & 500 & 0.46 & 0.44 & 0.50 & 0.49 & 0.00 & 1.93 \\ 
    CF-RF & 500 & 0.46 & 0.44 & 0.28 & 0.28 & 0.00 & 1.09 \\ 
    CF-XGBoost & 500 & 0.46 & 0.46 & 0.27 & 0.27 & 0.00 & 1.04 \\ 
    $\ell_1$-LR & 500 & 0.46 & 0.46 & 0.26 & 0.26 & 0.00 & 1.02 \\ 
    LR & 500 & 0.46 & 0.46 & 0.29 & 0.28 & 0.00 & 1.10 \\ 
    MARS & 500 & 0.46 & 0.44 & 0.30 & 0.30 & 0.00 & 1.16 \\ 
    RF & 500 & 0.46 & 0.51 & 0.24 & 0.24 & 0.00 & 0.93 \\ 
    Unadjusted & 500 & 0.46 & 0.44 & 0.26 & 0.26 & 0.00 & 1.00 \\ 
    XGBoost & 500 & 0.46 & 0.47 & 0.26 & 0.25 & 0.00 & 0.98 \\ \addlinespace 
    CF-MARS & 1500 & 0.46 & 0.87 & 0.28 & 0.27 & 0.00 & 1.04 \\ 
    CF-RF & 1500 & 0.46 & 0.88 & 0.27 & 0.27 & 0.00 & 1.03 \\ 
    CF-XGBoost & 1500 & 0.46 & 0.88 & 0.28 & 0.27 & 0.00 & 1.06 \\ 
    $\ell_1$-LR & 1500 & 0.46 & 0.88 & 0.27 & 0.26 & 0.00 & 1.00 \\ 
    LR & 1500 & 0.46 & 0.89 & 0.26 & 0.25 & 0.00 & 0.98 \\ 
    MARS & 1500 & 0.46 & 0.88 & 0.27 & 0.26 & 0.00 & 1.03 \\ 
    RF & 1500 & 0.46 & 0.91 & 0.26 & 0.26 & 0.00 & 0.98 \\ 
    Unadjusted & 1500 & 0.46 & 0.88 & 0.27 & 0.26 & 0.00 & 1.00 \\ 
    XGBoost & 1500 & 0.46 & 0.89 & 0.27 & 0.26 & 0.00 & 1.01 \\
    \bottomrule
  \end{tabular}
\end{table}

\subsection{Results for simulations with null treatment effect and where the
  covariates are not prognostic of the outcome}

\begin{table}[H]
  \centering
  \caption{Simulation results for the $\rmst$ of the time to
    intubation or death at day 14 under null treatment effect and covariates
    with no prognostic power.}
  \label{tab:rmst:nn}
  \begin{tabular}{lccccccc}
    \toprule
    Estimator & $n$ & Effect size & P(Reject H$_0$) & $n\times\mse$ & $n\times \var$ & |Bias| & Rel. eff.\\ \midrule
    CF-MARS & 100 & 0.00 & 0.11 & 142.00 & 135.82 & 0.25 & 1.62 \\ 
    CF-RF & 100 & 0.00 & 0.05 & 103.14 & 103.11 & 0.02 & 1.17 \\ 
    CF-XGBoost & 100 & 0.00 & 0.05 &  95.76 &  95.76 & 0.00 & 1.09 \\ 
    $\ell_1$-LR & 100 & 0.00 & 0.06 &  87.21 &  87.13 & 0.03 & 0.99 \\ 
    LR & 100 & 0.00 & 0.09 &  99.98 &  99.95 & 0.02 & 1.14 \\ 
    MARS & 100 & 0.00 & 0.08 &  97.67 &  97.24 & 0.07 & 1.11 \\ 
    RF & 100 & 0.00 & 0.31 &  70.72 &  70.68 & 0.02 & 0.80 \\ 
    Unadjusted & 100 & 0.00 & 0.06 &  87.92 &  87.85 & 0.03 & 1.00 \\ 
    XGBoost & 100 & 0.00 & 0.06 &  87.83 &  87.82 & 0.01 & 1.00 \\ \addlinespace 
    CF-MARS & 500 & 0.00 & 0.06 &  91.82 &  91.53 & 0.02 & 1.08 \\ 
    CF-RF & 500 & 0.00 & 0.05 &  98.39 &  98.38 & 0.00 & 1.16 \\ 
    CF-XGBoost & 500 & 0.00 & 0.05 &  97.91 &  97.91 & 0.00 & 1.15 \\ 
    $\ell_1$-LR & 500 & 0.00 & 0.05 &  87.33 &  87.32 & 0.01 & 1.03 \\ 
    LR & 500 & 0.00 & 0.06 &  92.74 &  92.74 & 0.00 & 1.09 \\ 
    MARS & 500 & 0.00 & 0.05 &  88.10 &  88.10 & 0.00 & 1.04 \\ 
    RF & 500 & 0.00 & 0.17 &  86.34 &  86.34 & 0.00 & 1.01 \\ 
    Unadjusted & 500 & 0.00 & 0.05 &  85.12 &  85.10 & 0.01 & 1.00 \\ 
    XGBoost & 500 & 0.00 & 0.06 &  86.89 &  86.89 & 0.00 & 1.02 \\ \addlinespace 
    CF-MARS & 1500 & 0.00 & 0.05 &  85.92 &  85.89 & 0.00 & 1.00 \\ 
    CF-RF & 1500 & 0.00 & 0.05 &  94.58 &  94.57 & 0.00 & 1.10 \\ 
    CF-XGBoost & 1500 & 0.00 & 0.04 &  93.27 &  93.26 & 0.00 & 1.09 \\ 
    $\ell_1$-LR & 1500 & 0.00 & 0.05 &  85.62 &  85.61 & 0.00 & 1.00 \\ 
    LR & 1500 & 0.00 & 0.06 &  90.04 &  90.01 & 0.00 & 1.05 \\ 
    MARS & 1500 & 0.00 & 0.05 &  89.32 &  89.32 & 0.00 & 1.04 \\ 
    RF & 1500 & 0.00 & 0.11 &  89.12 &  89.12 & 0.00 & 1.04 \\ 
    Unadjusted & 1500 & 0.00 & 0.05 &  85.87 &  85.83 & 0.00 & 1.00 \\ 
    XGBoost & 1500 & 0.00 & 0.07 &  89.61 &  89.59 & 0.00 & 1.04 \\
    \bottomrule
  \end{tabular}
\end{table}

\begin{table}[H]
  \centering
  \caption{Simulation results for the $\rd$ of the time to
    intubation or death at day 7 under null treatment effect and covariates
    with no prognostic power.}
  \label{tab:rd:nn}
  \begin{tabular}{lccccccc}
    \toprule
    Estimator & $n$ & Effect size & P(Reject H$_0$) & $n\times\mse$ & $n\times \var$ & |Bias| & Rel. eff.\\ \midrule
    CF-MARS & 100 & 0.00 & 0.10 & 1.11 & 1.07 & 0.02 & 1.52 \\ 
    CF-RF & 100 & 0.00 & 0.05 & 0.85 & 0.85 & 0.00 & 1.17 \\ 
    CF-XGBoost & 100 & 0.00 & 0.05 & 0.80 & 0.80 & 0.00 & 1.09 \\ 
    $\ell_1$-LR & 100 & 0.00 & 0.06 & 0.72 & 0.72 & 0.00 & 0.99 \\ 
    LR & 100 & 0.00 & 0.08 & 0.82 & 0.82 & 0.00 & 1.12 \\ 
    MARS & 100 & 0.00 & 0.07 & 0.79 & 0.79 & 0.00 & 1.09 \\ 
    RF & 100 & 0.00 & 0.27 & 0.66 & 0.66 & 0.00 & 0.90 \\ 
    Unadjusted & 100 & 0.00 & 0.06 & 0.73 & 0.73 & 0.00 & 1.00 \\ 
    XGBoost & 100 & 0.00 & 0.06 & 0.73 & 0.73 & 0.00 & 1.00 \\ \addlinespace 
    CF-MARS & 500 & 0.00 & 0.05 & 0.76 & 0.76 & 0.00 & 1.06 \\ 
    CF-RF & 500 & 0.00 & 0.05 & 0.82 & 0.82 & 0.00 & 1.13 \\ 
    CF-XGBoost & 500 & 0.00 & 0.05 & 0.80 & 0.80 & 0.00 & 1.11 \\ 
    $\ell_1$-LR & 500 & 0.00 & 0.05 & 0.72 & 0.72 & 0.00 & 1.00 \\ 
    LR & 500 & 0.00 & 0.06 & 0.78 & 0.78 & 0.00 & 1.08 \\ 
    MARS & 500 & 0.00 & 0.05 & 0.74 & 0.74 & 0.00 & 1.03 \\ 
    RF & 500 & 0.00 & 0.16 & 0.73 & 0.73 & 0.00 & 1.02 \\ 
    Unadjusted & 500 & 0.00 & 0.05 & 0.72 & 0.72 & 0.00 & 1.00 \\ 
    XGBoost & 500 & 0.00 & 0.06 & 0.72 & 0.72 & 0.00 & 1.00 \\ \addlinespace 
    CF-MARS & 1500 & 0.00 & 0.05 & 0.71 & 0.71 & 0.00 & 0.99 \\ 
    CF-RF & 1500 & 0.00 & 0.05 & 0.79 & 0.79 & 0.00 & 1.09 \\ 
    CF-XGBoost & 1500 & 0.00 & 0.04 & 0.77 & 0.77 & 0.00 & 1.07 \\ 
    $\ell_1$-LR & 1500 & 0.00 & 0.05 & 0.72 & 0.72 & 0.00 & 0.99 \\ 
    LR & 1500 & 0.00 & 0.06 & 0.74 & 0.74 & 0.00 & 1.03 \\ 
    MARS & 1500 & 0.00 & 0.05 & 0.74 & 0.74 & 0.00 & 1.03 \\ 
    RF & 1500 & 0.00 & 0.10 & 0.74 & 0.74 & 0.00 & 1.03 \\ 
    Unadjusted & 1500 & 0.00 & 0.05 & 0.72 & 0.72 & 0.00 & 1.00 \\ 
    XGBoost & 1500 & 0.00 & 0.06 & 0.75 & 0.75 & 0.00 & 1.05 \\
    \bottomrule
  \end{tabular}
\end{table}

\begin{table}[H]
  \centering
  \caption{Simulation results for the $\lor$ of the modified WHO scale
    at day 14 under null treatment effect and covariates with no
    prognostic power.}
  \label{tab:lor:nn}
  \begin{tabular}{lccccccc}
    \toprule
    Estimator & $n$ & Effect size & P(Reject H$_0$) & $n\times\mse$ & $n\times \var$ & |Bias| & Rel. eff.\\ \midrule
    CF-MARS & 100 & 0.00 & 0.07 & 38.32 & 38.30 & 0.01 & 0.99 \\ 
    CF-RF & 100 & 0.00 & 0.06 & 20.08 & 20.07 & 0.01 & 0.52 \\ 
    CF-XGBoost & 100 & 0.00 & 0.07 & 21.21 & 21.21 & 0.00 & 0.55 \\ 
    $\ell_1$-LR & 100 & 0.00 & 0.09 & 22.15 & 22.15 & 0.01 & 0.57 \\ 
    LR & 100 & 0.00 & 0.08 & 96.02 & 96.02 & 0.00 & 2.49 \\ 
    MARS & 100 & 0.00 & 0.07 & 28.91 & 28.90 & 0.01 & 0.75 \\ 
    RF & 100 & 0.00 & 0.11 & 16.61 & 16.60 & 0.01 & 0.43 \\ 
    Unadjusted & 100 & 0.00 & 0.06 & 38.60 & 38.60 & 0.00 & 1.00 \\ 
    XGBoost & 100 & 0.00 & 0.10 & 34.50 & 34.48 & 0.01 & 0.89 \\ \addlinespace 
    CF-MARS & 500 & 0.00 & 0.05 & 16.92 & 16.91 & 0.00 & 1.01 \\ 
    CF-RF & 500 & 0.00 & 0.05 & 17.27 & 17.27 & 0.00 & 1.03 \\ 
    CF-XGBoost & 500 & 0.00 & 0.06 & 17.75 & 17.74 & 0.00 & 1.06 \\ 
    $\ell_1$-LR & 500 & 0.00 & 0.06 & 16.45 & 16.45 & 0.00 & 0.98 \\ 
    LR & 500 & 0.00 & 0.06 & 17.28 & 17.28 & 0.00 & 1.03 \\ 
    MARS & 500 & 0.00 & 0.06 & 17.26 & 17.26 & 0.00 & 1.03 \\ 
    RF & 500 & 0.00 & 0.09 & 15.20 & 15.20 & 0.00 & 0.91 \\ 
    Unadjusted & 500 & 0.00 & 0.05 & 16.74 & 16.74 & 0.00 & 1.00 \\ 
    XGBoost & 500 & 0.00 & 0.07 & 16.54 & 16.54 & 0.00 & 0.99 \\ \addlinespace 
    CF-MARS & 1500 & 0.00 & 0.05 & 16.60 & 16.60 & 0.00 & 1.02 \\ 
    CF-RF & 1500 & 0.00 & 0.06 & 17.37 & 17.37 & 0.00 & 1.07 \\ 
    CF-XGBoost & 1500 & 0.00 & 0.06 & 17.11 & 17.11 & 0.00 & 1.05 \\ 
    $\ell_1$-LR & 1500 & 0.00 & 0.05 & 16.33 & 16.28 & 0.01 & 1.00 \\ 
    LR & 1500 & 0.00 & 0.05 & 16.61 & 16.61 & 0.00 & 1.02 \\ 
    MARS & 1500 & 0.00 & 0.05 & 16.25 & 16.24 & 0.00 & 1.00 \\ 
    RF & 1500 & 0.00 & 0.08 & 16.21 & 16.21 & 0.00 & 0.99 \\ 
    Unadjusted & 1500 & 0.00 & 0.05 & 16.30 & 16.29 & 0.00 & 1.00 \\ 
    XGBoost & 1500 & 0.00 & 0.06 & 16.30 & 16.29 & 0.00 & 1.00 \\
    \bottomrule
  \end{tabular}
\end{table}

\begin{table}[H]
  \centering
  \caption{Simulation results for the $\mw$ of the modified WHO scale
    at day 14 under null treatment effect and covariates with no
    prognostic power.}
  \label{tab:mw:nn}
  \begin{tabular}{lccccccc}
    \toprule
    Estimator & $n$ & Effect size & P(Reject H$_0$) & $n\times\mse$ & $n\times \var$ & |Bias| & Rel. eff.\\ \midrule
    CF-MARS & 100 & 0.50 & 0.08 & 0.42 & 0.42 & 0.00 & 1.54 \\ 
    CF-RF & 100 & 0.50 & 0.06 & 0.28 & 0.28 & 0.00 & 1.02 \\ 
    CF-XGBoost & 100 & 0.50 & 0.07 & 0.29 & 0.29 & 0.00 & 1.07 \\ 
    $\ell_1$-LR & 100 & 0.50 & 0.07 & 0.27 & 0.27 & 0.00 & 0.99 \\ 
    LR & 100 & 0.50 & 0.47 & 2.09 & 2.09 & 0.00 & 7.57 \\ 
    MARS & 100 & 0.50 & 0.08 & 0.36 & 0.36 & 0.00 & 1.32 \\ 
    RF & 100 & 0.50 & 0.11 & 0.22 & 0.22 & 0.00 & 0.79 \\ 
    Unadjusted & 100 & 0.50 & 0.06 & 0.28 & 0.28 & 0.00 & 1.00 \\ 
    XGBoost & 100 & 0.50 & 0.09 & 0.27 & 0.27 & 0.00 & 0.97 \\ \addlinespace 
    CF-MARS & 500 & 0.50 & 0.05 & 0.28 & 0.28 & 0.00 & 1.04 \\ 
    CF-RF & 500 & 0.50 & 0.06 & 0.28 & 0.28 & 0.00 & 1.04 \\ 
    CF-XGBoost & 500 & 0.50 & 0.06 & 0.29 & 0.29 & 0.00 & 1.07 \\ 
    $\ell_1$-LR & 500 & 0.50 & 0.05 & 0.26 & 0.26 & 0.00 & 0.99 \\ 
    LR & 500 & 0.50 & 0.08 & 0.72 & 0.72 & 0.00 & 2.69 \\ 
    MARS & 500 & 0.50 & 0.06 & 0.28 & 0.28 & 0.00 & 1.05 \\ 
    RF & 500 & 0.50 & 0.09 & 0.24 & 0.24 & 0.00 & 0.91 \\ 
    Unadjusted & 500 & 0.50 & 0.05 & 0.27 & 0.27 & 0.00 & 1.00 \\ 
    XGBoost & 500 & 0.50 & 0.07 & 0.26 & 0.26 & 0.00 & 0.99 \\ \addlinespace 
    CF-MARS & 1500 & 0.50 & 0.05 & 0.27 & 0.27 & 0.00 & 1.03 \\ 
    CF-RF & 1500 & 0.50 & 0.06 & 0.28 & 0.28 & 0.00 & 1.08 \\ 
    CF-XGBoost & 1500 & 0.50 & 0.06 & 0.28 & 0.28 & 0.00 & 1.06 \\ 
    $\ell_1$-LR & 1500 & 0.50 & 0.05 & 0.26 & 0.26 & 0.00 & 1.02 \\ 
    LR & 1500 & 0.50 & 0.05 & 0.27 & 0.27 & 0.00 & 1.03 \\ 
    MARS & 1500 & 0.50 & 0.05 & 0.26 & 0.26 & 0.00 & 1.00 \\ 
    RF & 1500 & 0.50 & 0.08 & 0.26 & 0.26 & 0.00 & 1.00 \\ 
    Unadjusted & 1500 & 0.50 & 0.05 & 0.26 & 0.26 & 0.00 & 1.00 \\ 
    XGBoost & 1500 & 0.50 & 0.06 & 0.27 & 0.27 & 0.00 & 1.02 \\
    \bottomrule
  \end{tabular}
\end{table}

\bibliographystyle{plainnat}
\bibliography{references}